\documentclass[12pt]{article}

\usepackage{epsf,epsfig}
\usepackage{graphicx}
\usepackage{amsfonts,amssymb,amsmath,rotating} 
\usepackage{fancyhdr,cite}
\usepackage[it]{caption}
\usepackage{citesort}

\newcommand{\degree}{\ensuremath{^\circ}}
\newcommand{\PDE}{PD\nobreakdash-E\,}
\newcommand{\rV}{$r$\nobreakdash-$V$\,}
\newcommand{\mean}{\ensuremath{\mathrm{mean}}}

\newcommand{\dminsfull}{\ensuremath{\left\{ d_x^0 \right\}_x}}
\newcommand{\dmaxsfull}{\ensuremath{\left\{ d_x^1 \right\}_x}}
\newcommand{\dmins}{\ensuremath{(\mathrm{min \ slopes})}}
\newcommand{\dmaxs}{\ensuremath{(\mathrm{max \ slopes})}}

\newcommand{\tmem}[1]{{\em #1\/}}

\begin{document}

\title{Estimating degrees of freedom in motor
systems}\author{Robert H. Clewley, John
M. Guckenheimer, \\and Francisco J. Valero-Cuevas}
\maketitle

\begin{abstract}

Studies of the degrees of freedom or ``synergies" in musculoskeletal
systems rely critically on algorithms to estimate the ``dimension" of
kinematic or neural data. Linear algorithms such as principal
component analysis (PCA) are used almost exclusively for this purpose.
However, biological systems tend to possess nonlinearities and operate
at multiple spatial and temporal scales so that the set of reachable
system states typically does not lie close to a single linear
subspace. We compare the performance of PCA to two alternative
nonlinear algorithms (Isomap and our novel pointwise dimension
estimation (\PDE)) using synthetic and motion capture data from a
robotic arm with known kinematic dimensions, as well as motion capture
data from human hands.  We find that consideration of the spectral
properties of the singular value decomposition in PCA can lead to more
accurate dimension estimates than the dominant practice of using a
fixed variance capture threshold. We investigate methods for
identifying a single integer dimension using PCA and Isomap. In
contrast, \PDE provides a range of estimates of fractal dimension.
This helps to identify heterogeneous geometric structure of data sets
such as unions of manifolds of differing dimensions, to which Isomap
is less sensitive.  Contrary to common opinion regarding fractal
dimension methods, \PDE yielded reasonable results with reasonable
amounts of data. We conclude that it is necessary and feasible to
complement PCA with other methods that take into consideration the
nonlinear properties of biological systems for a more robust
estimation of their degrees of freedom.

\end{abstract}

\section{Introduction\label{Introduction}}

The ability to use sensor data to objectively quantify the number of
active or controlled skeletal degrees of freedom (DOFs) during natural
behavior is central to the study of neural control of musculoskeletal
redundancy. A long standing problem in the study of neuromuscular
systems is whether and how the nervous system uses the numerous DOFs
provided by the neuro-musculo-skeletal system. For example, several
studies have sought to determine whether the nervous system couples
the mechanical DOFs of the hand to simplify the control of hand
shaping for grasp or sign language~\cite{PCAhand,WeissFlanders}. Other
important problems are the estimation of dimension of the neural
controller from electro\-myo\-graphic signals~\cite{dAvella} or
extra\-cellular neural recordings from the brain~\cite{Schwartz_etal}.
Theories of motor learning also address problems of dimension
estimation by proposing that the acquisition of complex tasks
progresses by initially ``freezing" some skeletal degrees of freedom
and gradually releasing them as the nervous systems is able to
incorporate them into the motor task~\cite{NewellVaillancourt}.

\subsection{Algorithmic methods to estimate the dimension of data}

This paper discusses algorithmic methods that measure the dimension in
state space occupied by observed dynamical behaviours. Our goal is to
compare and contrast the performance of today's mathematically
related, but distinct, definitions of dimension and varied approaches
to estimating the dimension of a dynamical system from sampled
data. We focus on comparing the performance of two established
algorithms---principal component analysis (PCA)~\cite{PCA} and
Isomap~\cite{Isomap}--- with a new algorithm that estimates pointwise
dimension (\PDE).

PCA, linear regression and {\em multi-dimensional scaling}
(MDS)~\cite{MDS} are linear methods that test whether a data set lies
close to a linear subspace, in which case the coordinates from this
subspace can be used to parameterize the data. However, these methods
do not determine whether the data may lie on a lower dimensional set
within the subspace. Indeed, a single arm rotating relative to the
body in a plane produces a motion capture data set that lies along a
circle, The circle is a one dimensional set because it can be
parameterized by a single coordinate (e.g., an angle), but the circle
does not lie close to a one dimensional linear subspace. Linear
methods suggest that two coordinates are most appropriate in this
example, where these coordinates describe the plane in which the
circular motion takes place. This is an example where linear methods
do not suffice in determining the dimension even of a simple geometric
object underlying the motion of a simple kinematic system.

{\em Isomap}~\cite{Isomap}, {\tmem{local linear embedding}}
(LLE)~\cite{LLE}, and {\tmem{Laplacian}} or {\tmem{Hessian
eigenmaps}}~\cite{LapEmaps,HessEmaps} are methods that have been 
developed within the setting of machine learning 
and dimension reduction to find coordinate systems for nonlinear
manifolds. They include procedures for discovering the dimension
of data sets that lie on smooth Riemannian manifolds. Isomap 
seeks a set of global coordinates for this manifold via singular 
value decomposition of a matrix of interpoint distances of the 
data.

In our application to biomechanics the underlying structure of the
biomechanical system governing, say, locomotion or manipulation, may
not be representable as motion in a smooth manifold. The structure
might instead decompose as the union of submanifolds having different
dimension for different phases in the gait cycle (e.g., swing versus
double support) or grasp acquisition versus manipulation.  We need
techniques that can identify the appropriate decomposition and
estimate the dimension of the different phases of the task.  This
paper shows that \PDE aids in the process of exploring this kind of
complex geometric structure in data sets.

{\em Pointwise dimension} is a quantity assigned to probability
densities or measures that are defined on metric spaces. Like Isomap,
algorithms for computing pointwise dimension are based upon analysis
of the distances between pairs of data points.  However, the way in
which this information is used is quite different in the two
methods. Algorithms for estimating the pointwise dimension of
attractors of dynamical systems were developed in the 1980's and
applied to many different empirical data sets.  Much of this work used
a technique called ``delay embedding'' to manufacture multidimensional
data from a single (one-dimensional) time series. To use delay
embedding effectively the time step between successive observations
and the number of successive observations to use in the embedding have
to balanced to account for the sensitive dependence of solutions to
initial conditions, the dimension of the attractor and the
independence of observations made at each time step. There was a
consenus that prohibitive amounts of data were required by the methods
for accurately estimating the pointwise dimension of high dimensional
attractors~\cite{GWSP}. This paper revisits the numerical estimation
of pointwise dimension in the context of motion capture data, where
the emphasis is upon estimating the dimension of sets that are already
embedded in high-dimensional Euclidean spaces.

\subsection{Goal and approach}

We compare and assess the ability of
\PDE, PCA, and Isomap methods to estimate the dimension
of data sets that are representative of motion capture.  The methods
are applied to synthetic data sets generated from simple geometric
objects and from a numerical simulation of a robot arm. As
representative examples of empirical data, we study the kinematic DOFs
of a robotic arm and the kinematics of the human fingers using motion
capture data. The motion capture data consist of the spatial locations
of identified markers placed on the surface of the systems as they
assume a large set of postures representative of their workspace.

Our results suggest that estimation of pointwise dimension is a
promising tool for the analysis of motion capture data, and in
particular we estimate bounds on the DOFs involved in hand kinematics.
We suggest avenues to expand the mathematical theory underlying our
method, and propose additional empirical testing necessary to explore
its most effective use.

\section{Methods\label{methods}}

\subsection{Pointwise fractal dimension\label{ptwisedim}}

Algorithms for estimating the {\tmem{pointwise}} and
{\em correlation} dimensions of data sets were developed in the
1980's within the context of assigning dimensions to attractors of
dynamical systems
\cite{DoyneF83,Guck84_dimest,Young_dim}. These algorithms
are framed in the setting of {\em measures} or {\em probability
densities} in a metric space, and assume that the data whose dimension
is being determined is distributed like independent samples of the
measure.  They do not make use of the temporal structure of
trajectories and can be applied to arbitrary data sets that give
discrete approximations to a probability measure $\mu$. In practice,
this means that the $\mu$-measure of a set $S$ (we call this the
{\em volume} of $S$) can be approximated by the {\em proportion}
of data points that lie in $S$.  Methods such as PCA, Isomap, and LLE
presume that input data represent samples from a geometric set with
the structure of a Riemannian manifold. In contrast, pointwise
dimension and the algorithms used to estimate it make sense for sets
that are made up of a union of manifolds having different dimension
and for large classes of fractal sets supporting a suitable measure.
Biological systems are likely candidates for pointwise dimension
approaches given that their structure, function and control operate at
different spatial and temporal scales and in different modes
simultaneously (e.g., skin
vs.\ fat vs.\ muscle vs.\ bone motions; reflex vs.\ voluntary control of
movement; etc.).

The pointwise dimension of $x \in S$ is defined by measuring the
growth rate of balls of radius $r$ centered at $x$ as a function of
$r$. Denoting the balls by $B_x(r)$, the dimension
$d_{\mu}(x)$ of $\mu$ at $x$ is
\begin{equation}
d_{\mu}(x) = \lim_{r \to 0} \frac{\log(\mu(B_x(r))}{\log(r)}.  \label{ptwisedim_def}
\end{equation}
\noindent This limit may not exist and it may not be the same for all
points of $S$. When it does exist, it reflects a power law scaling in
which the volume of balls is proportional to $r^d$.  The pointwise
dimension of measures has been studied in the context
of dynamical systems~\cite{Young_dim,Barreira_dim}.
\footnote{Young~\cite{Young_dim} established the existence and
measurability of the pointwise dimension for so-called SRB measures of
dynamical systems. Barreira~\cite{Barreira_dim} and others have
extended the definition of pointwise dimension to some non-ergodic
measures.  For such measures $\mu$, the dimension is $\mu$-almost
everywhere the same, and this is defined to be the pointwise dimension
of $\mu$.}  Here, we adopt a pragmatic approach in the context of
experimentally-obtained data sets.

Calculation of the pointwise dimension of a data set $S$ with $N$
points can be implemented efficiently with sorting algorithms. Given a
reference point $x$, the distances between $x$ and all other $N-1$
points $y$ in the data set are calculated and sorted.  If $r_k$ is the
$k^{\mathrm{th}}$ distance in the sorted list, then we estimate
$\mu(B_x(r_k)) = k/(N-1)$. A dimension estimate of $S$ for reference
point $x$ is the asymptotic slope of $\log(\mu(B_x(r)))$ vs $\log(r)$
as $r \to 0$. If there is a good linear regression fit of
$\log(\mu(B_x(r_k)))$ to $\log(r_k)$, then the slope of this line is
taken as an estimate for $d_{\mu}(x)$. However, for reasons discussed
in our definition of the \PDE algorithm below, we cannot expect the
data in this log-log plot to be well fit by a line over the entire
range of observed values of $r$. Thus, choosing the range of $r$ over
which to fit the log-log plot is subjective. We discuss our
implementation choices below.

\subsection{Dimension estimation algorithms\label{dimest}}

\subsubsection{PCA\label{PCA_methods}}

Assume that we have a data set of $N$ observations in a $D$
dimensional Euclidean {\em data space} with $N>D$.  In the case of our
motion capture data, $k$ markers are placed upon an object and
analysis of video recordings produces the spatial locations of these
markers, yielding a data space of dimension $D = 3k$. PCA is a linear
method for testing whether the data lie close to a linear subspace $U
\subset \mathbb{R}^D$ whose dimension $d$ is smaller than $D$. The
first step of PCA is to normalize the data and assemble data vectors
into a $D \times N$ matrix $A$.  The next step is to calculate the
singular value decomposition $A = W \Sigma V^t$ where $W$ and
$V$ are orthogonal $D \times D $ and $N \times D$ matrices and
$\Sigma$ is a $D \times D $ diagonal matrix of singular values $\xi_i$,
ordered by decreasing magnitude. Projection onto the subspaces
$U_l$ spanned by the first $l$ columns of $W$ minimizes the
mean squared residual ($L^2$ norm) of the original (normalized)
data among projections onto $l$ dimensional subspaces of the data
space and maximizes the variance of the projected data.

We define the cumulative norm of the first $i$ singular values as
$s(i) = \sqrt{(\xi_1^2 + \ldots + \xi_i^2)}$ for $i=1, \ldots, D$, and denote
its maximum value $\hat{s} \equiv s(D)$.  From this we define the
fraction of variance explained up to dimension $i$ as $\sigma(i) =
s(i) / \hat{s}$, and the corresponding residual (fraction of variance
unexplained) as $\rho(i) = \sqrt{ 1-\sigma(i)^2 }$. $\sigma(i)$ is a
monotonically increasing function of $i$, while $\rho(i)$ is
monotonically decreasing.

Estimating the dimension of the data set from PCA requires a criterion
for choosing a minimal $l$ for which the projected data is an
acceptable ``reduction.'' A frequent choice (e.g.,~\cite{PCAhand}) for
this criterion is to fix a variance capture threshold, given by the
algorithmic parameter $\tau < 1$ such that $\sigma(l)>\tau$.

A second choice that is seldom used in the biomechanics literature is
to select a value of $l$ for which there is a ``knee'' in a linear-log
graph of the residuals $\rho(i)$: i.e., the quantities $\rho(i) -
\rho(i+1)$ are substantially larger for $i<l$ than for $i > l=1$.
This method is less sensitive to noise and is better tuned to the
scaling properties of an individual data set. For PCA, we implement this
criterion by computing the second differences of $\log(\rho(i))$ and
determine when these are larger than a threshold given by an
algorithmic parameter $\gamma$. Where there are one or more
consecutive second differences larger than $\gamma$, we declare there
to be a knee at the local maximum of the second differences. We found
that the value $\gamma=0.1$ caused the algorithm to select knee
positions that corresponded best to positions that we judged by eye.

\subsubsection{Isomap\label{isomap_methods}}

The Isomap (isometric mapping) algorithm seeks to reconstruct the
Riemannian metric on a submanifold of the data space and find global
coordinates that preserve this metric. One assumes that the data set
of $N$ points in $\mathbb{R}^D$ lies on a submanifold and that it
samples this manifold densely enough that the Euclidean distance
between near neighbours in the data set approximates distance along
the manifold. Neighbourhoods consisting of these near neighbours are
encoded in a ``neighbourhood graph'' with $N$ vertices, one for each
data point. Vertices are connected by undirected edges in this graph
in one of two ways: (1) vertex $v_i$ is connected to its $K$ nearest
neighbours in the data set, or (2) $v_i$ is connected to vertices $v_j$
for which the corresponding distances satisfy $|| x_i - x_j || <
\varepsilon$. Geodesic distance between $x_i$ and $x_j$ is then
estimated by minimizing $$\sum_{k=1}^l || x_{k+1} - x_k ||$$ among
chains of points $x_i = x_0, \ldots, x_l = x_j$ which come from paths
in the neighbourhood graph. The resulting distances are represented in
the matrix $G(i,j)$. The neighbourhood graph may be disconnected, in
which case the data is partitioned by components of the neighbourhood
graph for further analysis. Here we retain only the component with the
largest number of points.

Isomap then uses the classical multi-dimensional scaling method on the
matrix $G$, producing a singular value decomposition.  We estimate the
dimension of the data set with Isomap using a method similar to that
described above for PCA, but the residual variance is defined
differently. The fraction of variance captured $\sigma(l)$ for Isomap
measures how much the matrix of $L^2$ distances between the first $l$
singular vectors of the multi-dimensional scaling decomposition
covaries with $G$. The residual variance is then $1-\sigma(l)$, which
need not be a monotonically decreasing function of $l$.  We search for
either a minimum (when the function is non-monotonic) or a point of
maximum curvature (when the function is monotonic). We use the same
criterion for detecting a knee in a linear plot of the residual
variance using $\gamma=1$.

In all our tests with the Isomap algorithm we selected evenly-spaced
``landmark'' points in the data at a sample rate of 1 in every 10
regular data points. As recommended by Tenenbaum et al.~\cite{Isomap},
this results in many more landmark points than the expected dimension
of the data and also many fewer than $N$.

Isomap can be run using either a selection of the neighbourhood radius
$\varepsilon$ or the number of nearest neighbours $K$ as the principal
parameter.  As outlined by Tenenbaum et al.~\cite{Isomap2}, we made a
trade-off between two cost functions in order to select these Isomap
parameters appropriately: the fraction of the variance in geodesic
distance estimates not accounted for in the Euclidean embedding, and
the fraction of points not included in the largest connected component
of the neighbourhood graph, and thus not included in the Euclidean
embedding of that component.

If $K$ or $\varepsilon$ are chosen large enough that all interpoint
distances are retained, then the identity map gives the manifold
metric of the sampled data and Isomap will detect only the dimension
of a linear subspace containing the data.  Similarly, when these parameters
are chosen small enough so that only a very few interpoint distances
are retained, the graph of neighbouring points becomes disconnected
or the estimation of geodesic distances along a manifold
are no longer accurate. When these instances arise in the Results we
will simply indicate that Isomap failed to produce a dimension
estimate.

We will also discuss the use of pointwise dimension estimation results
in guiding initial choices for these parameters.

\subsubsection{Pointwise Dimension Estimation\label{PD-E}}

We now describe a new empirical method for estimating the pointwise
dimension of a data set of $N$ points in $\mathbb{R}^D$ which we refer
to as {\tmem{Pointwise Dimension Estimation}} (\PDE).  The heart of
the method is the scaling relationship between the volume $V$ of a
ball and its radius $r$: $V \sim r^d$ in dimension $d$. We assume that
a data set is a discrete approximation of a probability measure
$\mu$. Pragmatically, this means that the proportion of the points of
the data set that lie in a set $S$ is assumed to be approximately
$\mu(S)$. More specifically, we interpret the proportion of data
points within distance $r$ as an estimate for the volume of the ball
$B_x(r)$ of radius r centered at $x$.  If $x$ is in our data set and
we have tabulated the matrix of interpoint distances, then sorting the
distances from $x$ is an efficient way of determining the function
$m_x(r) = \mu(B_x(r))$. We directly get the inverse of $m_x$ via the
observation that the distance from $x$ to its $k^{\mathrm{th}}$
nearest neighbour gives the value of $r$ for which $m_x(r) = k/(N-1)$.
The scaling relationship $m_x(r) \sim r^d$ is equivalent to
$\log(m_x(r)) = d \log(r) + c$ for some constant $c$. Thus the \PDE
method is based upon the following steps:
\begin{enumerate}
\item
Select a {\em reference point} $x$ from the data set.
\item
Compute the $N-1$ distances $r_k$ from the other points of the data 
the reference point $x$.
\item
Sort the distances $r_k$.
\item 
Construct the log-log plot of $\log(k)$ vs. $\log(r_k)$.
\item
Estimate the slope of this log-log plot. We call these plots \rV {\em curves}.
\end{enumerate}

\begin{figure}[tbp]
\centerline{\ \includegraphics[scale=0.55]{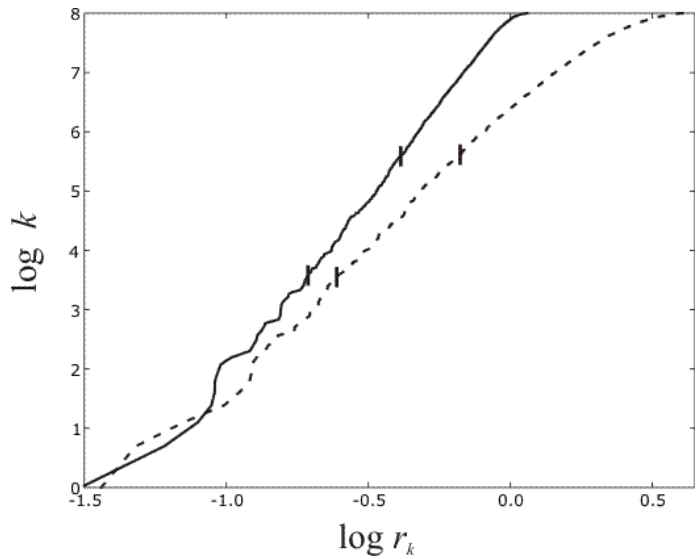} }
\caption[]{At the heart of the \PDE method is the determination of
slopes in a log-log plot of the index of data points sorted by
distance from a reference point as a function of that distance
(referred to as an \rV curve). The slopes indicate esimates of an
exponent $d$ in a power-law relationship between the radius $r$ and
the volume $V$ (approximated by the number of points $k$ in the ball
$r$): i.e., $V \sim r^d$. $d$ is an estimate of the dimension of the
set of points. This figure shows two such graphs for points randomly
distributed in a 6-dimensional ball. The full line is the graph for a
reference point at the centre of the ball, and the dotted line is the
graph for a reference point with coordinates $(0.57, -0.14, -0.12,
0.37, 0.66, 0.03)^t$.  The marker lines on the graphs indicate a range
within which \PDE selects the neighbourhood's inner and outer radii
and computes the slope of the secant defined by the two radii on this
graph. The slopes for the central reference point and the distal
reference point are 6.1 and 5.0, respectively. (Reference points near
to the boundary of data sets are more likely to generate \rV curves
with lower slopes and thus cause the dimension of the sets to be
underestimated.)
\label{log-log}}
\end{figure}

If the \rV curves are linear and have the same slope for all reference
points, then this common slope is the pointwise dimension of the data
set.  However, there are inevitable statistical fluctuations and other
sources of deviations of the \rV curves from linear
functions that occur in this procedure. Figure~\ref{log-log} shows two
\rV curves for a data set of 3,000 independent random
samples from the uniform Lebesgue measure in a six dimensional unit
ball. One curve has its reference point at the center of the ball,
while the second curve has a randomly chosen reference point.  Note
that the curves have substantial fluctuations from a straight line at
small values of $\log(r)$ and that the second curve deviates from a
straight line also at large values of $\log(r)$. This example points
to the need for additional analysis to extract good estimates of the
pointwise dimension of $\mu$ from the \rV curves.
 
Among the sources of variability in the slopes of the \rV curves are
the following:
\begin{enumerate}
\item
Sampling errors that reflect the difference between the data set and the 
probability measure $\mu$ it is assumed to approximate.
\item
Noise in the data is expected to yield measures that are always $D$
dimensional, but the amplitude of the noise is likely to make the
support of the measure ``thin'' in some directions.
\item
The pointwise dimension of the measure $\mu$ may not exist, and even
if it does, the slope of the log-log plots may have asymptotic slope
$d$ only for $\mu$-almost all reference points. This is a typical
situation for dynamical system attractors related to their
multifractal structure~\cite{Fractal_meas}.
\item
The ``shape'' of the dataset and $\mu$ affects the slope of the
log-log plot at larger distances from the reference point. As an
example, the slope of the log-log plot for uniform measure on a two
dimensional rectangle with side lengths $R_1 \ll R_2$ will be
approximately one for radii $R_1 < r < R_2$. In high dimensional balls
and rectangular solids, a very large proportion of the measure is
concentrated near the boundary of the set, leading to slopes of the
\rV curves that are substantially smaller than the dimension.
\end{enumerate} 

In the absence of firm mathematical foundations for estimating
pointwise dimension, we have pursued empirical tests on observational
and simulated data. We have experimented with techniques for selecting
a suitable ``scaling'' region of the \rV curves that exclude small
distances (smaller than $r_x^{min}$) subject to large sampling
fluctuations and noise, and large distances (larger than $r_x^{max}$)
where the global shape of the object plays a dominant role in
determining the relationship between volume and radius. We have also
experimented with ways of representing the statistical distribution of
slopes with the scaling regions of \rV curves.  We assume that a
random selection of a moderate number of reference points suffices to
approximate the distribution of these slopes for the measure $\mu$.
Unless otherwise stated we select $0.2N$ reference points.

Based on the above observations and assumptions we characterize the
varying slopes of the \rV curves in this paper in the following
way. For the \rV curve of each reference point $x$, we ignore the five
points closest to the reference point, and the furthest 30\%.  With
the remaining points, a range of secants are determined along the
curve. The lower positions of the secants are selected at uniform
intervals in the space of nearest neighbour indices, in steps of
$0.0005 N$ (rounded, if necessary). Thus, the lower point of contact
on the \rV curve for a secant with index $k$ is
$\mathrm{log}\,(k)$. The upper index is chosen to be $\Delta k$, where
we choose the constant $\Delta = 4$ based on observing the typical
scale of regions of near-constant slope on the curves. The upper point
of contact is therefore $\mathrm{log}\,(\Delta k)$. The range of $k$
indices for secants ends where $\Delta k$ becomes equal to or greater
than $0.7 N$.  The minimum and maximum of the slopes of the secants
are recorded for reference point $x$, and are denoted $d_x^0$ and
$d_x^1$, respectively.  We then calculate the minimum, maximum, mean,
median, and inter-quartile range of the sets $\dmins =
\dminsfull$ and $\dmaxs = \dmaxsfull$ defined over the range of
reference points $x$.

We plot representative \rV curves for reference points
corresponding to the extrema and the means of these sets. We also use
a scatter plot of all $(d_x^0, d_x^1)$ pairs as a function of $x$ to
characterize the distribution of slopes found. We highlight the points
in the scatter plot corresponding to the extrema and means using a
colour code. The full description of the graphical presentation of
these statistics is given in the caption to
Figure~\ref{6d_ball}.

We also considered an alternative approach for assigning slopes to \rV
curves based on linear regression. That approach produced estimates of
dimension within the range of the method described here.  The
determination of the slopes of secants requires fewer calculations
than attempting to fit straight line segments on the \rV curves using
linear regression.

\subsection{Computer generated synthetic data\label{randomdata_methods}}

We tested the \PDE algorithm on independent samples from measures of
known dimension. The test measures we used are Lebesgue measure on 6-
and 54-dimensional rectangular solids and balls.  (The dimension of the
space of motion capture marker data from our robot arm is 54, as there
are 18 reflective markers placed on the robot.) We analyzed
points uniformly distributed in a rectangular solid with sides of
unit length, and from one that has 4 sides one fifth of the length of
the remaining unit-length sides. This enables us to explore the fact that the
relative length scales of different directions in the data are an
issue for dimension estimation algorithms. We investigated sample sizes
between 2,000 and 8,000 points.

We also performed tests of the algorithms involving
the ``Swiss roll'' surface used as a benchmark by Tenenbaum et
al.~\cite{Isomap}.  The Swiss roll is a two dimensional surface coiled
smoothly in a three dimensional space.  We generated one-dimensional
curves on this surface parameterized by an angle $\theta $ (both
closed and open curves), from which we uniformly sampled 1,000 points.
Before rolling the two dimensional surface that contains the curve,
the $(x,y)$ coordinates for the $i^{\mathrm{th}}$ of $N$ sample points
were generated by the formulae $(0.5 \, \mathrm{cos}\,\theta + b_x
i/N, 0.5\, \mathrm{sin}\, \theta + b_y i/N)$. For the closed curves,
we set $(b_x, b_y) = (0,0)$, whereas for the open curves the values
used were $(1.0, 0.2)$. The transformation that rolled these points in
the plane into three dimensions is given by $(x,y) \mapsto \left(
(1.5-1.4x) \, \mathrm{cos}\, 4 \pi x, y, (1.5-1.4x) \, \mathrm{sin}\,
4 \pi x \right)$. The relative scales of the Swiss roll manifold are
set so that the curve extends approximately one quarter of the
distance in the $y$ direction as in the $x$ and $z$ directions.

\begin{figure}[tbp]
\centerline{\includegraphics[scale=0.7]{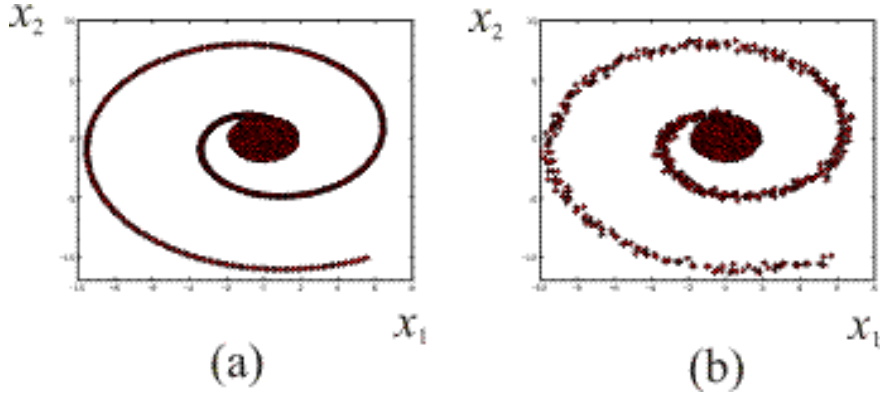} }
\caption[]{(a) and (b) A projection of points sampled randomly from a planar disc and points generated along a single spiral arm in the same plane, embedded in 5 dimensions. The disc consists of 2,000 points, and the spiral arm contains 1,000 points. In (b) Gaussian noise was added to all points.\label{disc_spiral}}
\end{figure}

In a third set of benchmark tests, synthetic data sets were generated
by sampling from a set that is a union of two manifolds of different
dimensions. This consisted of a planar disc region (with a radius of
1.8) and one or two spiral arms emanating from the disc in the same
plane. See Figure~\ref{disc_spiral}. The spirals were generated as
involutes of a circle having half the radius of the disc, using the
relationship $r^2 = \phi^2 +1$ in polar coordinates, yielding an
inter-arm distance of approximately twice the radius of the disc. The
total diameter of the data set in the plane was approximately
18. 2,000 points were sampled from the disc and 1,000 from the spiral
arm. These data were then embedded in a five-dimensional ambient
space. In some tests Gaussian noise was added to all five coordinates
with a standard deviation equal to approximately 5\% of the radius of
the disc. The range of the noisy data in directions orthogonal to the
disc and spiral is approximately $\pm 0.5$, which is approximately
30\% of the disc's radius.

\subsection{Motion capture data for robot arm\label{robot_methods}}

An AdeptSix 300 robot arm with six rotational joints was used to
produce motion capture data.  These data sets tested our analytical
techniques on a real mechanical system whose active DOFs are known
precisely. The motion of a robot arm is similar to a musculoskeletal
system, but exhibits less noise and has no passive DOFs.

\begin{figure}[tbp]
\centerline{\ \includegraphics[scale=0.55]{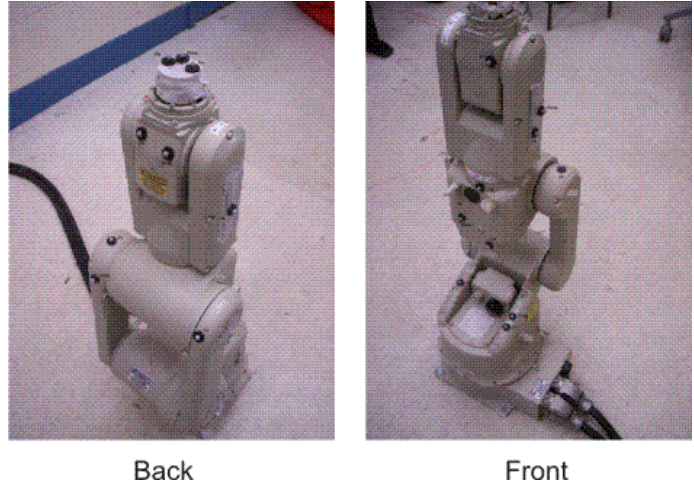} }
\caption[The ``home'' configuration of the AdeptSix 300 robot arm.]{The ``home'' configuration of the AdeptSix 300 robot arm, showing the reflective markers used for 3D motion capture.\label{robot1}}
\end{figure}

\begin{figure}[tbp]
\centerline{\ \includegraphics[scale=0.65]{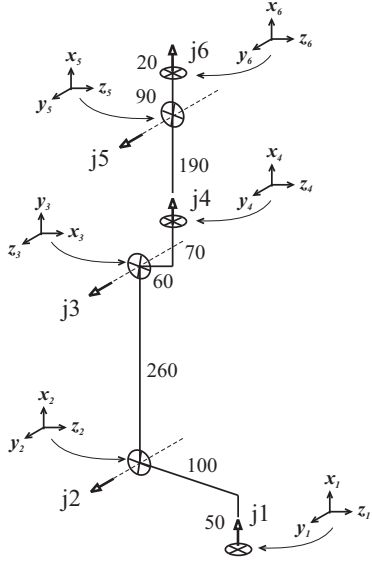} }
\caption[Schematic diagram of the robot arm.]{Schematic diagram of the physical dimensions of the robot arm (units are mm), including indication of joint axes and their local Euclidean coordinate frames (used for specifying marker positions).\label{robotdims}}
\end{figure}

The ``home'' configuration of the robot arm can be seen in
Figure~\ref{robot1}.  Figure~\ref{robotdims} is a schematic diagram of
the arm showing the local Euclidean coordinate frames defined around
each link.  The total length of the links is approximately 800\,mm.

Three reflective markers were attached around each joint (see
Figure~\ref{robot1}) for the purpose of tracking the robot's posture
by a 4-camera optical motion capture system manufactured by Vicon
(Vcams, Vicon Workstation). Table~\ref{markerpostable} provide details
of the reflective marker positions using the local
coordinate axes for the joints. Marker data were captured at a rate of
100\,Hz.  Only frames in which all markers were visible and properly
reconstructed were kept in the final data set. The mean calibration
residual of the Vicon marker reconstruction is less than 0.2\,mm. The
standard deviation in the reconstructed distance between two markers
on a rigid object is approximately 0.05\,mm.

\begin{table}[tbp]
\renewcommand{\arraystretch}{1.2}
\begin{center}
\begin{tabular}{|c||c|c|c|} \hline
Joint & Marker 1 & Marker 2 & Marker 3 \\ \hline \hline
1 & (70, -100, 20) & (64, 100, 0) & (58, 100, -20) \\ \hline
2 & (275, -120, 40) & (255, 110, 45) & (300, 110, 5) \\ \hline
3 & (-120, 30, -55) & (-60, -30, -55) & (30, 15, -50) \\ \hline
4 & (100, 75, 50) & (70, 75, 25) & (70, -85, -20) \\ \hline
5 & (10, -30, 65) & (20, 30, 60) & (10, 30, -55) \\ \hline
6 & (30, -35, 0) & (30, -20, -30) & (30, 32, 7) \\ \hline
\end{tabular}
\end{center}
\caption[]{Approximate marker positions from joint axes in local Euclidean joint coordinates (units are mm).\label{markerpostable}}
\end{table}

Two experiments were performed with the AdeptSix 300. In the first,
the robot arm was programmed to move to a succession of joint angles
in a random walk that cyclically varies a single joint angle at
each step. Conservative limits were placed on each angle choice to
prevent the robot from hitting either the floor or itself during
motion, and to keep all the markers within the range of visibility of
the cameras.  The constraints used on each joint were $\pm30\degree,
\pm15\degree, \pm15\degree, \pm45\degree,
\pm15\degree, \pm120\degree,$ respectively. The speed of robot
movement was selected to expedite the trial times but the motion did
not exhibit extraneous oscillations. The transition time between target
postures was approximately $1/3\,s$.

An initial set of random angle displacements were selected. These were
added to the angles associated with the robot's default posture, and
individually reselected when any of the angle limits were passed. This
generated the first target posture of the robot. The next position
was selected by randomly choosing the angular displacement of joint
\#1. The joint to be updated was cycled through the six joints for
determining each subsequent posture. The distribution of points in
joint space produced by this protocol is random, but may not be uniform.

\begin{figure}[tbp]
\centerline{\includegraphics[scale=0.55]{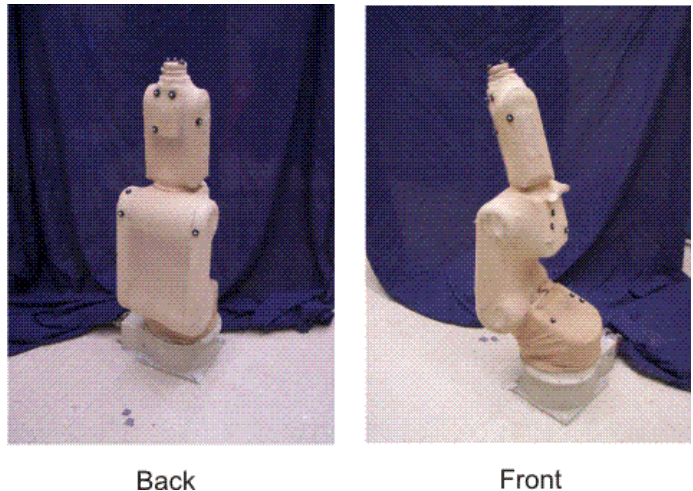} }
\caption[The sheathed robot arm and marker placement.]
{The sheathed robot arm and marker placement.\label{robot2}}
\end{figure}

The joint angle targets chosen in the first experiment were recorded
so that exactly the same sequence of targets could be reproduced in a
second experiment. The second experiment differed by putting the
entire robot arm inside a tight elastic sheath (white hosiery,
attached by elastic bands around joints 2 and 4), and re-attaching the
markers in positions as close as feasible to their prior positions in
the nominal configuration (see Figure~\ref{robot2}). The sheath was
intended to provide a source of systematic noise in the measurement of
the robot's actual motion, in this case to mimic the effects of skin
in the reconstruction of animal skeletal motion using markers attached
to the skin.

2.5 hours of data were collected from each experiment, but this was
resampled at a rate of approximately 1 frame per three seconds,
resulting in a data set of approximately 4,000 points.

We did not low-pass filter the kinematic data in our method so that we
retained any correlated high-frequency components (``synergies'') of
motion along with actual noise.  It is a goal of our analytical
methods to separate sources of noise and correlated high-frequency
components.

\subsection{Virtual robot motion\label{virtualrobot_methods}}

We further tested our methods with synthetic data of a simulated robot
arm without an elastic sheath. We reconstructed the geometry of the
AdeptSix 300 robot in a kinematic chain model of the joints, placing
the same number of markers in approximately the same positions. The
model was positioned by setting the six joint angles, and the forward
kinematic transformation from angle space into Euclidean marker space
was performed to generate marker positions of the virtual robot.

We used two different methods to generate joint angles of the arm.
One method was the same random walk protocol used for the physical
robot.  The second was to use independent samples of a uniform measure
in joint space.  The same limits on the joint angles were used as for
the physical robot. The chosen angles were mapped into marker space to
obtain the data set.

The role of experimental measurement accuracy and noise can be
investigated in the dimension estimation algorithms using the virtual
robot.  Also, different experimental protocols for selecting postures
can be easily evaluated using simulations. The effect of limits on the
joint angle selection on the output of the \PDE algorithm can
be evaluated in the virtual domain because we can safely remove them
and allow the virtual arm's motion to be unconstrained by
self-intersections of the robot arms or intersections with the floor.
Additionally, there are issues as to the relative scales of the
marker-axis distances for the different joints, which we would like to
be able to vary in order to explore its effect on a dimension estimate
calculation. The virtual robot allows us to position the markers
arbitrarily around the axes.

\subsection{Constrained hand motion\label{hand_methods}}

We used the Vicon 3D motion capture system to record time-series
kinematic data of two subjects asked to perform three tasks. The
subjects held their wrists in a fixed position while moving their
fingers. Five reflective markers were placed on each finger (one at
the fingertip and two between each of the joints), three on the thumb,
and four additional markers were placed on the back of the hand (a
total of 27 markers).

The first task was simultaneous ``random" movement of their fingers
close to the plane of the palm. The other two tasks were the
simulation of typing on a computer keyboard and the simulation of
manipulation of a track ball, both while keeping the wrist in a fixed
position. These tasks were performed for approximately 20 minutes in
four 5 minute segments, and the resulting data sets combined and
resampled to select 3 frames per second.  This resulted in final data
sets containing approximately 8,000 points.

\section{Results\label{results}}

\subsection{Computer generated synthetic data\label{randomdata_results}}

We expect that Isomap and PCA will detect that dimension reduction is
inappropriate for data sampled randomly from Lebesgue measure on a
sphere or rectangular solid in $\mathbb{R}^D$.  We tested data sets of
independent random samples from a 6-dimensional unit ball, using
sample sizes of 2,000, 4,000, and 8,000.  PCA at the 90\% variance
capture threshold determined $d=6$. Also, graphs of PCA residuals and
Isomap residual variances showed no ``knees,'' indicating that using
PCA or Isomap in this manner predicts that dimension reduction is not
appropriate, as expected.  However, PCA at the 80\% variance capture
threshold determined $d=5$, implying that for a sufficiently low
threshold this use of PCA incorrectly predicts that dimension
reduction is appropriate.

\begin{figure}[tbp]
\centerline{\includegraphics[scale=0.9]{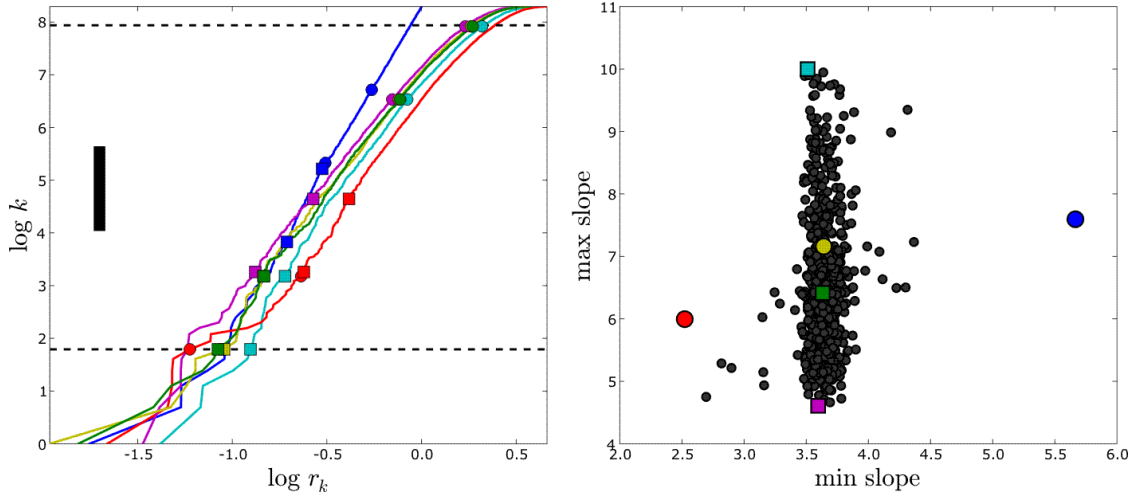} }
\caption[]{\PDE analysis of a uniformly-random distribution of sample points from a 6-dimensional unit ball. The left panel shows \rV
curves for the colour-coded points in the scatter plot in the right
panel. These highlighted points indicate the minimum, maximum and mean
of the minimum and maximum slopes. Round markers indicate statistics
relating to the minimum slopes, square markers relate to the maximum
slopes.  The colour coding is as follows: red = minimum of $\dmins$;
yellow = mean of $\dmins$; blue = max of $\dmins$; magenta = min of
$\dmaxs$; green = mean of $\dmaxs$; cyan = max of $\dmaxs$. We define $\dmins =
\dminsfull$ and $\dmaxs = \dmaxsfull$. The secant
end points estimating the minimum and maximum slopes of all the \rV
curves plotted are indicated by round and square markers,
respectively.  The dotted lines mark the closest and furthest nearest
neighbours considered in the estimation of slopes. The solid bar
indicates the size of the free parameter $\mathrm{log}\,\Delta$.  The
broad distribution of maximum slopes in the scatter plot can be
largely attributed to the distance of the associated reference point $x$
from the centre of the ball.\label{6d_ball}}
\end{figure}

\begin{figure}[tbp]
\centerline{\ \includegraphics[scale=0.75]{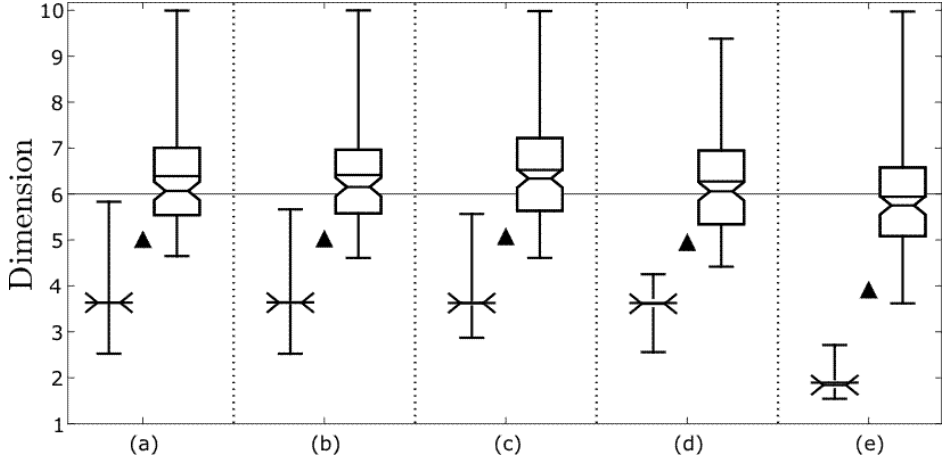} }
\caption[]{PD-E dimension estimates for a uniformly-random distribution of sample points from a 6-dimensional solid: (a)~a unit ball with $N=8000$, (b)~a unit ball with $N=4000$, (c)~a unit ball with $N=2000$, (d)~a unit cube with $N=2000$, and (e)~a rectangular solid having 2 sides of unit length and 4 sides of length 0.2, with $N=2000$.
For each solid, two box-and-whisker plots are shown. The left plot indicates the distribution of $\dmins$, the right indicates that of $\dmaxs$. In each case, the whiskers mark the extent of the data (from minimum value to maximum), the boxes mark the inter-quartile range. The notches indicate the median value. The horizontal line in the box indicates the mean value. When the inter-quartile range is very small, the top and bottom of the box is not drawn for the sake of clarity: the range is still apparent by the distance from the median to the beginning of each whisker. The black triangle between each pair of box plots indicates the mean of the data from $\dmins$ and $\dmaxs$ taken together. The thin horizontal line across each panel indicates the known dimension $D$ of the data set analyzed.
Panels (a)--(c) show the relative insensitivity of \PDE results on the number of points in the data set. Panel~(d) shows that the data set's non-smooth boundary does not significantly affect the results, compared to the smooth boundary in~(c). Panel~(e) shows that the relative scales of the solid's side lengths distort the distribution of \PDE dimension estimates.
\label{ball6_results}}
\end{figure}

Figures~\ref{6d_ball} and~\ref{ball6_results}(a)--(c) summarize the
results of the \PDE analysis on the balls. The dimension was
accurately estimated by the median of the maximum slopes of \rV curves.

\begin{figure}[tbp]
\centerline{\includegraphics[scale=0.95]{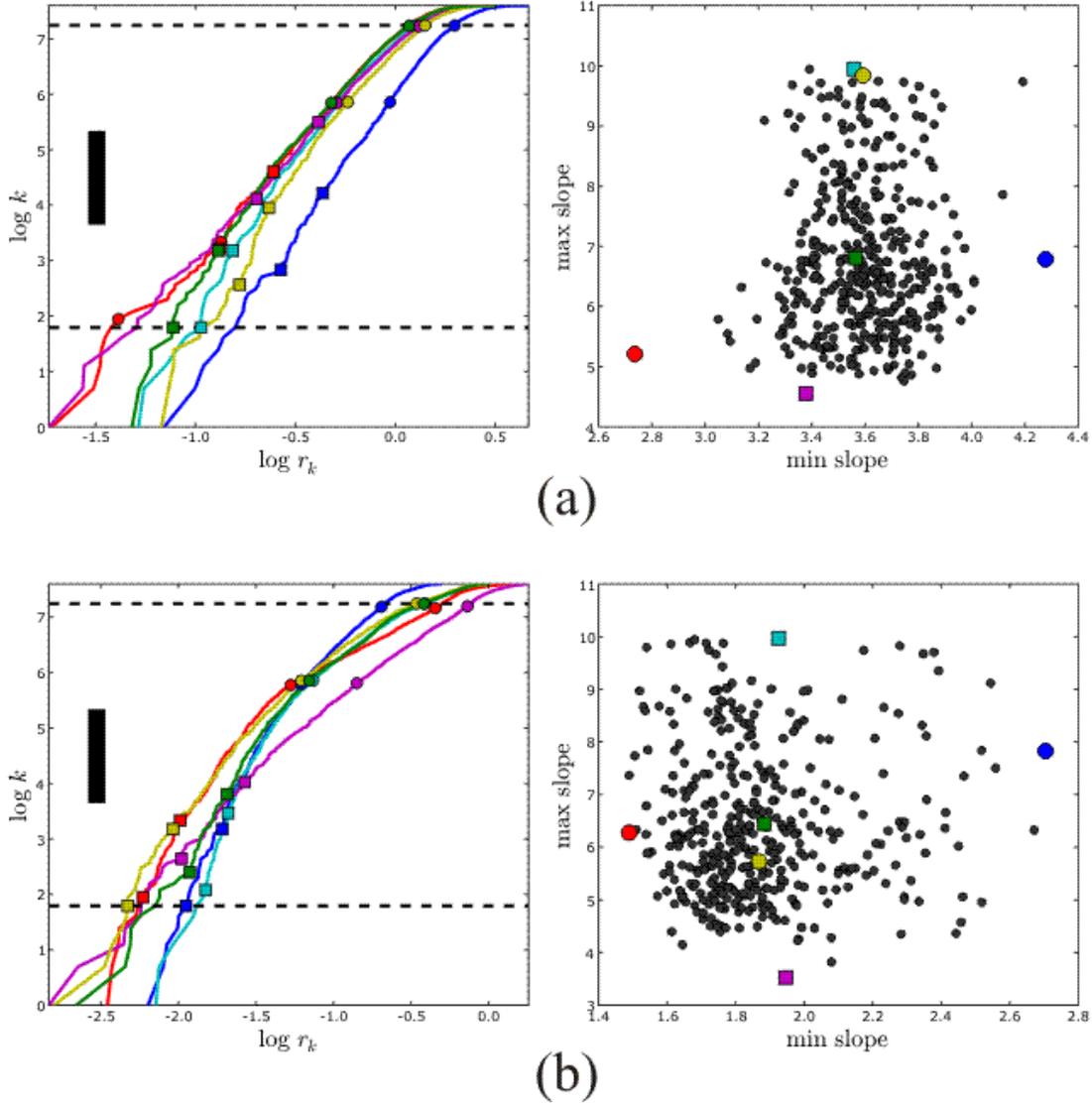} }
\caption[]{Output of the \PDE algorithm for the uniform random data sampled from a 6-dimensional rectangular solid with (a) equal side lengths, and (b) 4
sides one fifth the length of the other 2. In~(b) the slopes of the \rV curves can be seen to flatten out for $\log \: r_k > -1$, a result of balls $B_x(r)$ growing outside the solid in the shorter directions for lower values of $r$.
\label{random_rect_histos_6D}}
\end{figure}

We compare these results with an analysis of 2,000 points drawn
randomly from a 6-dimensional unit cube. The output of \PDE is
shown in Figure~\ref{random_rect_histos_6D}(a), and corresponding
statistics also appear in Figure~\ref{ball6_results}. (The results were
almost identical for $N=4000$.)  The PCA results at 80\% and 90\%
variance capture thresholds determined $d=5$ in both cases. The
distribution of maximum slopes $d_x^1$ is very similar between the
cube and the ball, but the minimum slopes $d_x^0$ are more widely
spread for the cube. From the \rV curves plotted, the minimum slopes
appear to be found at larger $r_x$, which is where we expect the most
distortion due to boundary effects of the set.  These effects are seen
even more strongly in Figure~\ref{random_rect_histos_6D}(b) for the
6-dimensional rectangular solid with four sides having one fifth the
length of the remaining two. Again, the minimum slopes are found
mostly at larger $r_x$ and their distribution is more
dispersed than that of the maximum slopes.

\begin{figure}[tbp]
\centerline{\includegraphics[scale=0.6]{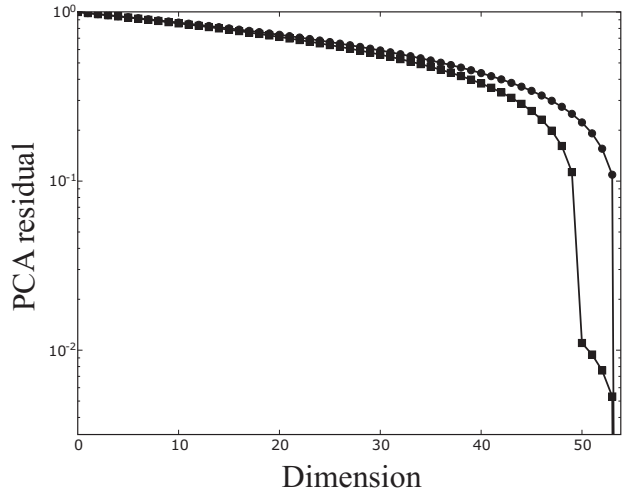} }
\caption[]{Linear-log plot of PCA residuals as a function of embedding dimension, for data set from 54-dimensional rectangular solid with equal side lengths (round markers) and with 4 sides one fifth the length of the other 50 (square markers). A knee is detected in the latter at $d=50$. The commonly-used criterion for choosing the estimated dimension $d$ in PCA uses a pre-selected ``variance capture'' threshold, which leads to sensitivity of the estimates to noise variance and generally an underestimation of dimension. In contrast, the detection of knees in this linear-log graph of the residuals performs more reasonably. Here, it correctly predicts that dimension reduction is not appropriate for the 54D solid with equal side lengths (no knee), although it incorrectly predicts that a mild reduction to $d=50$ is appropriate for the 54D solid with unequal side lengths.
\label{random_rect_PCA_54D}}
\end{figure}

Figure~\ref{random_rect_PCA_54D} shows the graphs of PCA residuals
for points sampled from rectangular solids with $D=54$.  For a rectangular
solid with unequal side lengths, a knee was detected in the graph of
PCA residuals at $d=50$ that reflects the decreased variance of the
data in four directions. At the 80\% and 90\%
variance capture levels PCA estimated the dimension of the
54-dimensional rectangular solid with equal (unequal) sides to be 42
and 48 (39 and 45), respectively.

\begin{figure}[tbp]
\centerline{\includegraphics[scale=0.9]{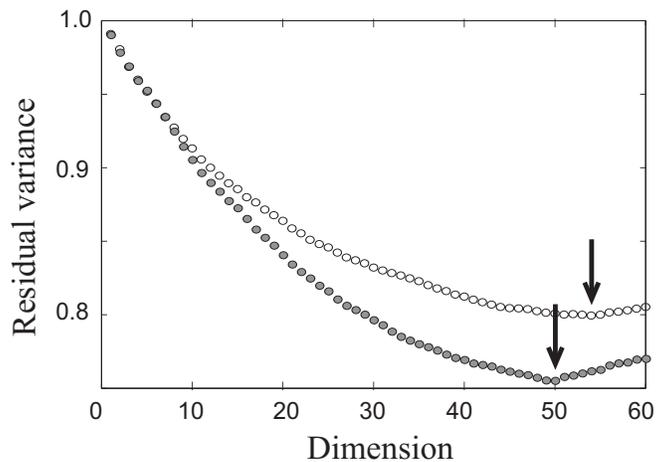} }
\caption[]{Plot of residual variance from Isomap embeddings ($K=50$) applied to data sampled from a 54-dimensional rectangular solid with equal
side lengths (open circles), and a 54-dimensional rectangular solid
with 4 sides one fifth the length of the other 50 (solid circles). The
arrows indicate the positions of the minima in the plots corresponding
to the best estimate of dimension of the data
set. The detection of knees in these graphs as an indication that dimension reduction
is appropriate correctly predicts that it is not appropriate for the 54D solid with equal side lengths, although it incorrectly predicts that a mild reduction to $d=50$ is appropriate for the 54D solid with unequal side lengths. Isomap's performance on these sets are essentially the same as that of PCA using the knees in the graphs of residuals (Figure~\ref{random_rect_PCA_54D}).
\label{random_isomap}}
\end{figure}

In order to apply Isomap we down-sampled these data sets by a factor
of one half, to fit the size limitations of the current implementation
of Isomap. Using $K=50$ we
obtained the distribution of residual variances as a function of
embedded dimension shown in Figure~\ref{random_isomap}. Similar
results were obtained for larger values of $K$, and values of
$\varepsilon>4$.  The knees of these distributions correspond to 54
for the rectangular solid with equal side lengths and 50 for the
solids with four shorter sides.

\PDE analysis of these 54-dimensional examples gives results that require
more complex interpretations.  As $N$, the number of sample points,
increases, the \rV curves show the volumes of balls intersected with
the region $R$ from which the samples are drawn. When the radius of
the ball exceeds the distance from the reference point to the boundary
of $R$, the slope of the \rV curve decreases. In high dimensions, most
of the measure of a sphere or rectangular solid is concentrated close
to its boundary since volume grows proportionally to $r^D$. Also, the
smallest interpoint distance that we expect to find in a data set of
$N$ points increases with dimension.  Thus we expect that the slopes
computed from the \rV plots will underestimate $D$.

\begin{figure}[tbp]
\centerline{\includegraphics[scale=0.95]{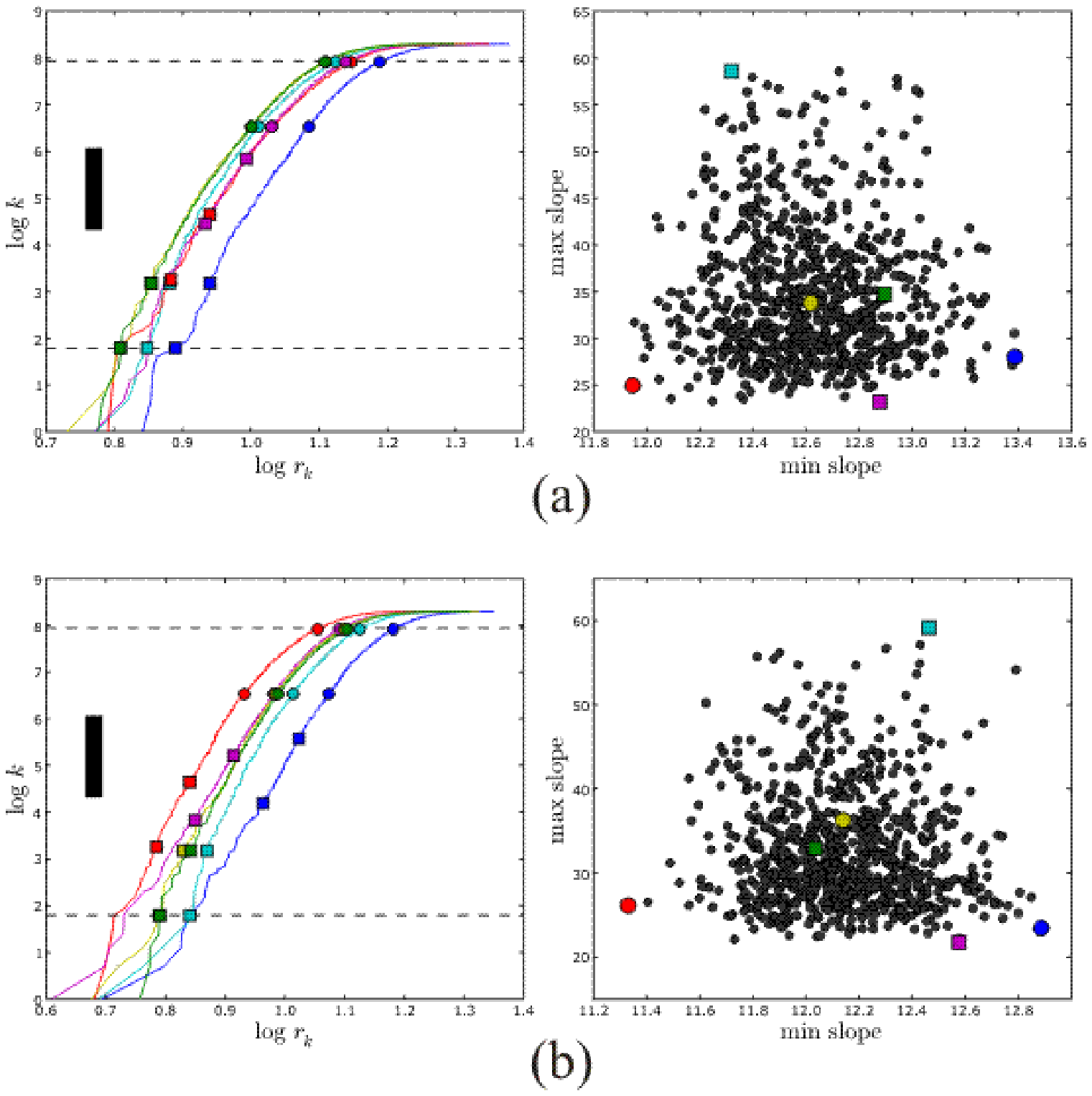} }
\caption[]{\PDE analysis for the uniform random data sampled from a
54-dimensional rectangular solid having (a)~equal side lengths, and
(b)~4 sides one fifth the length of the other 50. The flattening out of slopes in the
\rV curves for larger radii $r_k$ is less apparent for the solid with 50 long sides and
4 short sides in~(b), compared to the 6D solid in
Figure~\ref{random_rect_histos_6D} where there are only 2 long sides
compared to 4 short sides.
\label{random_rect_histos_54D}}
\end{figure}

\begin{figure}[tbp]
\centerline{\includegraphics[scale=0.65]{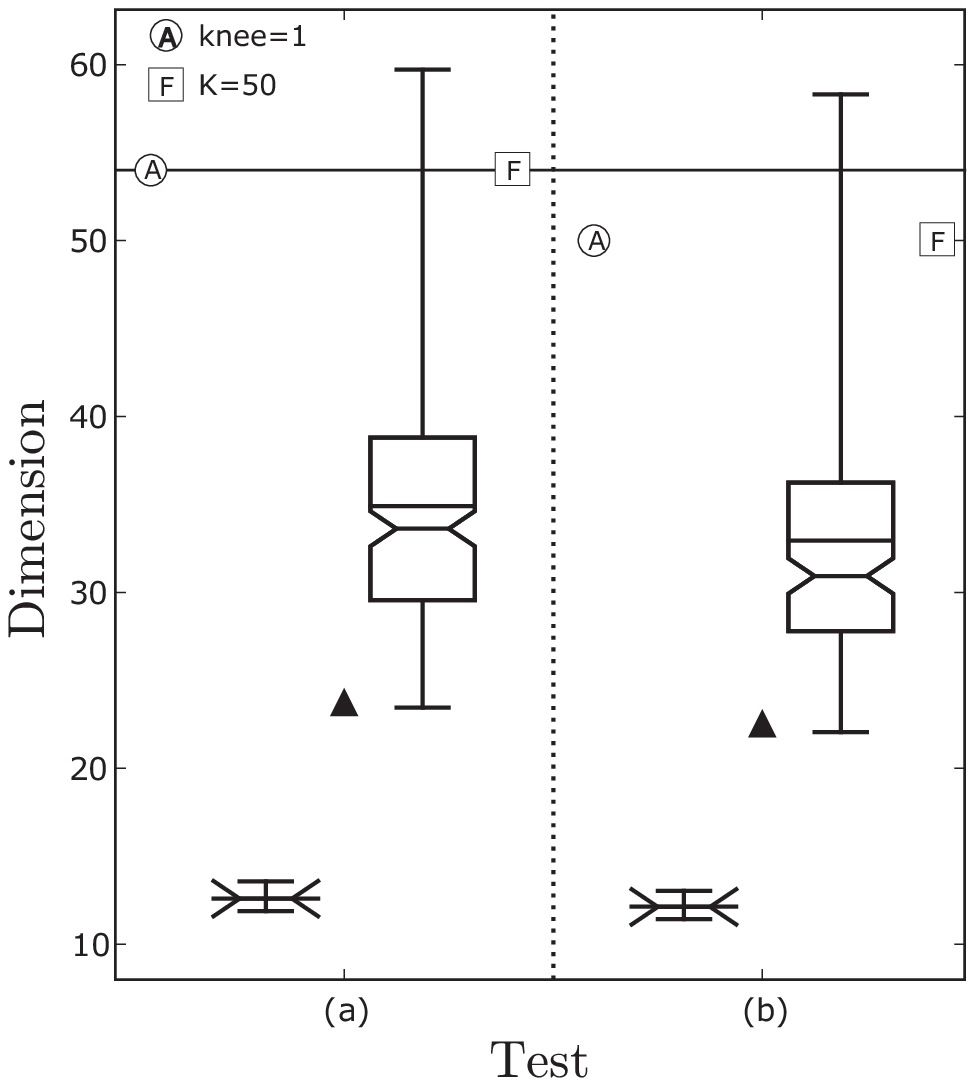} }
\caption[]{Dimension estimates of 54-dimensional rectangular solids having (a)~equal side lengths, and (b)~4 sides one fifth the length of the other 50.  A description of the box-and-whisker plots is given in Figure~\ref{ball6_results}.
The circular markers indicate the dimension estimation result for PCA using the first knee of the residual. The square markers indicate the dimension estimation results for Isomap using the parameter indicated.
For such high-dimensional sets \PDE does not accurately predict $D$ but still provides a reasonable upper bound on the dimension.
\label{rect54_results}}
\end{figure}

\begin{figure}[tbp]
\centerline{\includegraphics[scale=0.8]{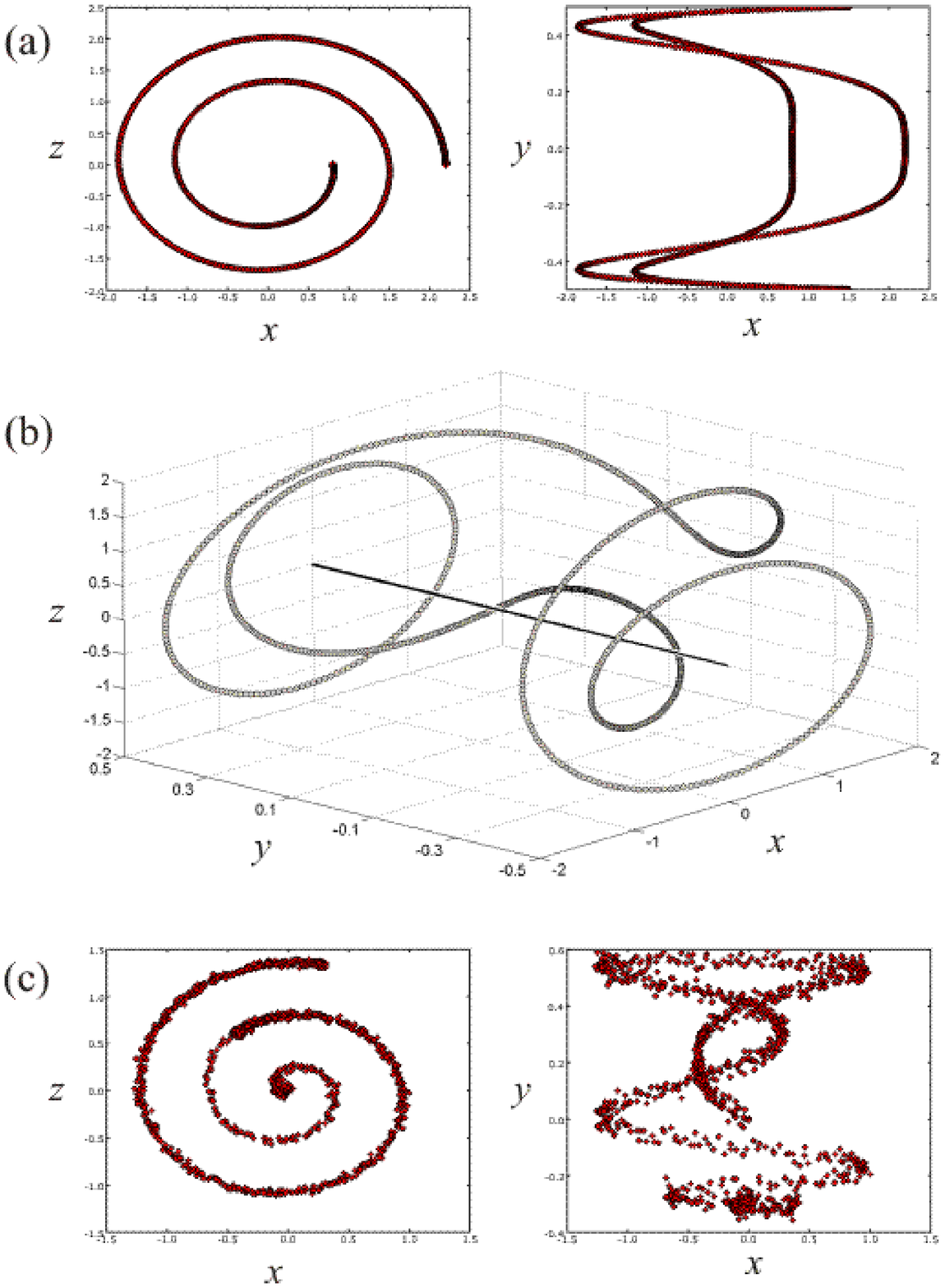} }
\caption[]{(a) A sample of 1,000 points from a closed curve embedded on a Swiss roll manifold, projected in to the $(x,z)$ and $(x,y)$ planes. (b)~3D projection of the curve in panel (a) showing the central axis of the roll along $(x,z) = (0,0)$. (c)~A noisy sample of 1,000 points from an open curve on a Swiss roll manifold, shown in the same projections as panel (a). Note that the $y$-axis scaling is one quarter that of the other axes.\label{swiss_trajs}}
\end{figure}

We tested the algorithms on a curve of points lying on a Swiss roll
manifold to assess their ability to estimate the dimension of a highly
nonlinear and low-dimensional manifold. In Figure~\ref{swiss_trajs}(a)
the points are evenly distributed on a closed curve and the sampling
is noiseless. In Figure~\ref{swiss_trajs}(c) the curve is open, and
Gaussian noise was added with zero mean and a standard deviation of 0.02
(corresponding to about 4\% of the spatial separation between the
layers of the rolled manifold).

\begin{figure}[tbp]
\centerline{\includegraphics[scale=0.95]{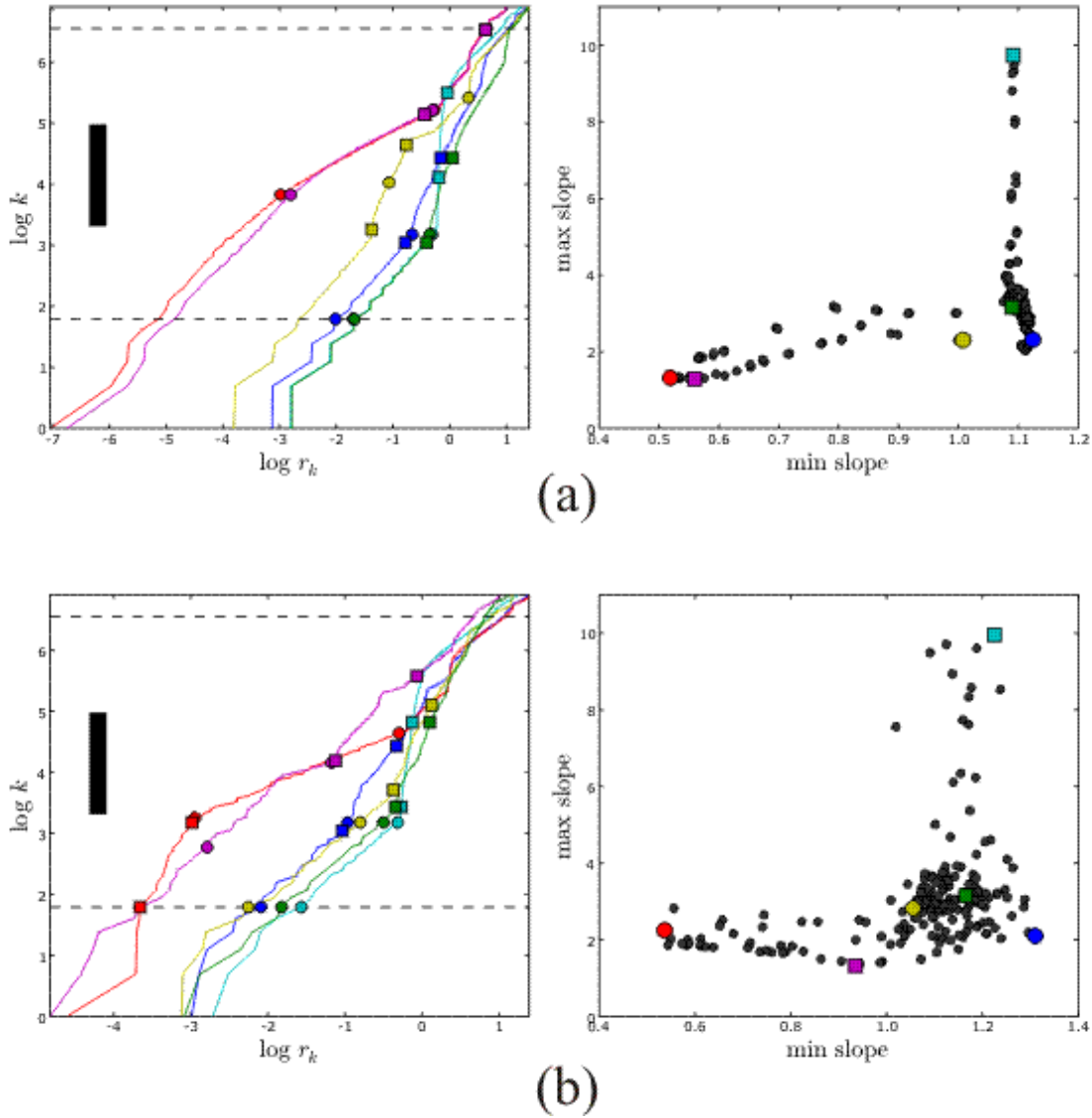} }
\caption[]{(a) \PDE analysis for the closed curve on a Swiss roll manifold. (b)~\PDE analysis for the open curve on a Swiss roll manifold with added Gaussian noise.
The minimum slopes of \rV curves occur most typically for small radii $r$, whereas the maximum slopes occur most typically for large radii. At the smaller scales (smaller $r$) the minimum slopes are clustered near the known dimension of the curve, $d_x^0=1$. At larger scales (larger $r$) the maximum slopes $d_x^1$ detect the dimension of the 3D ambient space in which the curve is embedded.\label{swiss_results}}
\end{figure}

\begin{figure}[tbp]
\centerline{\includegraphics[scale=0.75]{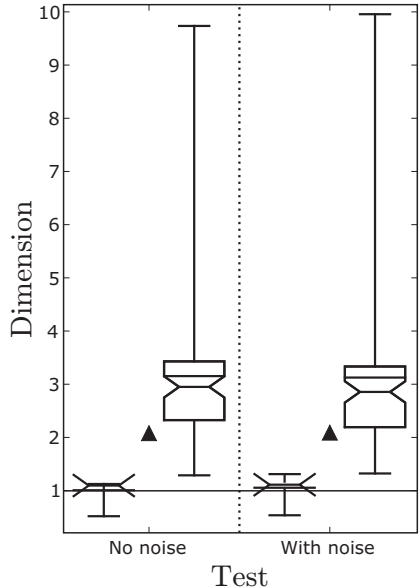} }
\caption[]{\PDE estimates for curves lying in the Swiss roll manifold. A description of the box-and-whisker plots is given in Figure~\ref{ball6_results}. The presence of noise does not substantially alter the dimension estimates.
\label{swiss_dims}}
\end{figure}

The results of the \PDE analysis of the closed curve data set are
shown in Figure~\ref{swiss_results}(a).  The statistics corresponding
to these are given in Figure~\ref{swiss_dims}. There is a clear
clustering in the scatter plot of minimum and maximum slopes that
corresponds to large and small radii in the \rV curves.  Balls of
small radius intersect a single branch of the curve and volumes
increase linearly with $r$. Balls of larger radius intersect an
increasing number of branches of the curve and display an increase in
volume that is faster than linear, reflected in the larger slope of
the \rV curves. For the open curve with added noise the output of the
algorithm shows some dispersal of the cluster of points associated
with $d_x^0 \approx 1.1$ along the axis of minimum slope
(Figure~\ref{swiss_results}(b)), but little dispersal of the cluster
associated with $d_x^1=3$ along the other axis. This is to be expected
because the amplitude of the noise is small compared to the global
extent of the curve in the three-dimensional space.
 
PCA at a threshold of 80\% and 90\% variance capture identified 2
dimensions for both the noise-free and noisy data sets.  In Isomap we
set the $\varepsilon$ parameter to be a length scale known to be much
smaller than the distance between the rolls of the surface
($\varepsilon = 0.08$).  The residuals for embedding in dimensions 1,
2, and 3 for the noise-free data were $1.4\times 10^{-7}$, $7.8\times
10^{-6}$, and $2.1\times 10^{-5}$, respectively, indicating a minimum
immediately at $d=1$.  The residuals for the noisy data were
$3.1\times 10 ^{-4}$, $1.9\times 10^{-4}$, $4.2 \times 10^{-4}$.
Therefore, Isomap estimates $d=1$ for the noise-free data and 2 for
the noisy data. However, if greater length scales are used for
$\varepsilon$ then the residual variance for $d=1$ becomes
approximately a magnitude larger than for $d>1$, meaning that Isomap
estimates $d=2$ in both cases.  Using the alternative control
parameter $K$, Isomap only detected $d=1$ for these data sets for $K$
between 5 and 15.

\begin{figure}[tbp]
\centerline{\includegraphics[scale=0.95]{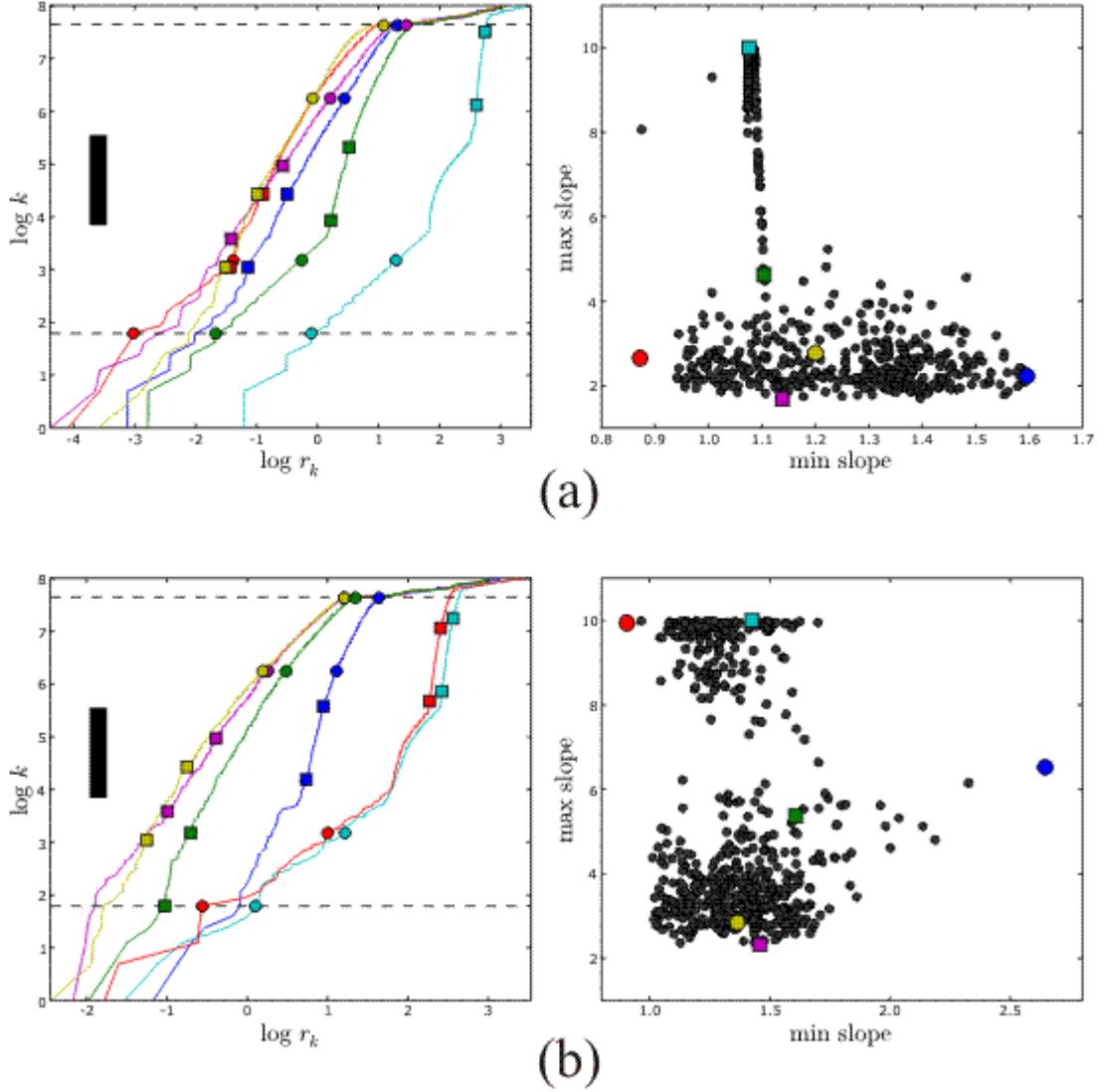} }
\caption[]{\PDE analysis for the data set sampled from the union of a disc and a co-planar spiral embedded in a 5-dimensional space. (a)~No noise added. (b)~Gaussian noise added with zero mean and a standard deviation of 0.1 in each of 5 dimensions. The clustering in the distribution of maximum slopes into two major concentrations is due to the data set being heterogeneously distributed in its ambient space (see text for details).\label{discspiral_results}}
\end{figure}

\begin{figure}[tbp]
\centerline{\includegraphics[scale=0.55]{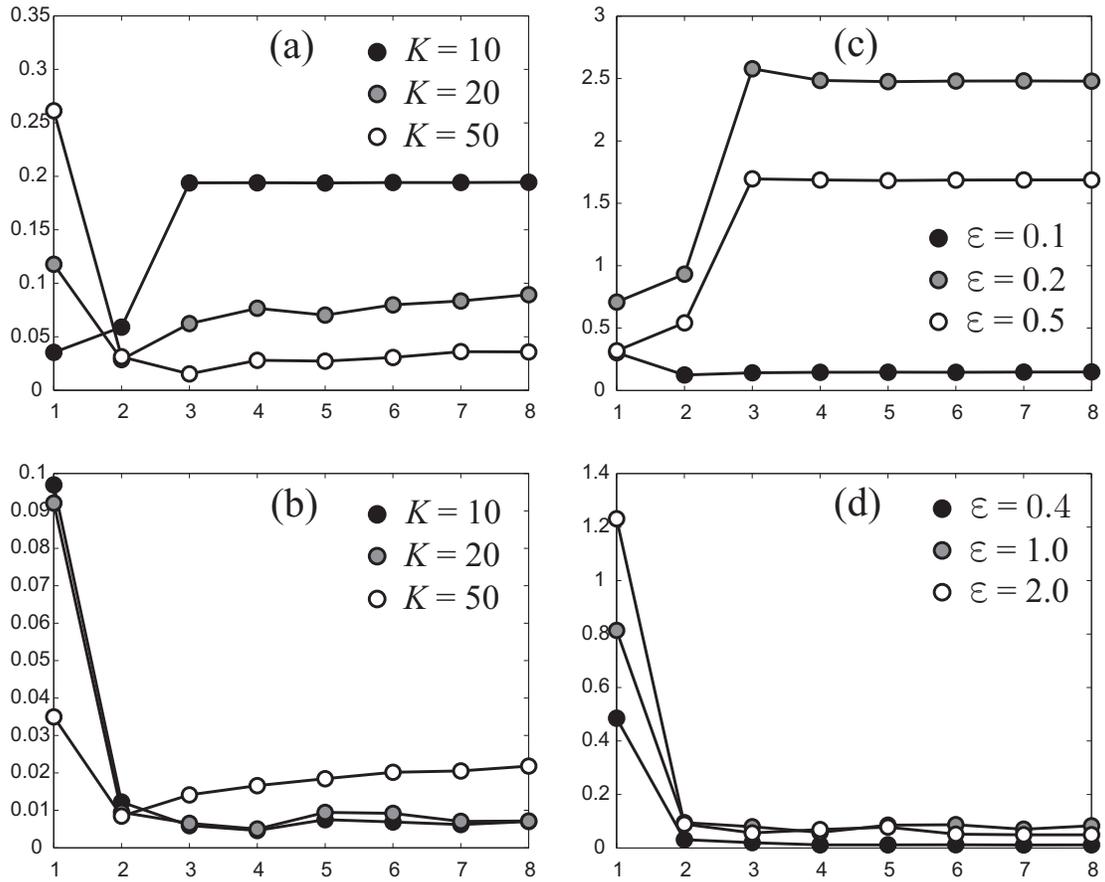} }
\caption[]{Graphs of Isomap residual variance versus embedding dimension $d$ for the data set sampled from the union of a disc and a co-planar spiral embedded in a 5-dimensional space. (a)~Control parameter $K$, no noise added. (b)~Control parameter $K$, noise added. (c)~Control parameter $\varepsilon$, no noise added. (d)~Control parameter $\varepsilon$, noise added. The vertical scales are not equal between parameter values,
but only the relative change of each graph of residual variance as a function of $d$ is important.
These graphs show that the estimate $d$ is more sensitive to the parameter value $K$ than to $\varepsilon$, as the knees or minima in the residual variances as a function of $d$ vary between 2 and 4 when $K$ is varied.
\label{isomap_discspiral}}
\end{figure}

\begin{figure}[tbp]
\centerline{\includegraphics[scale=0.73]{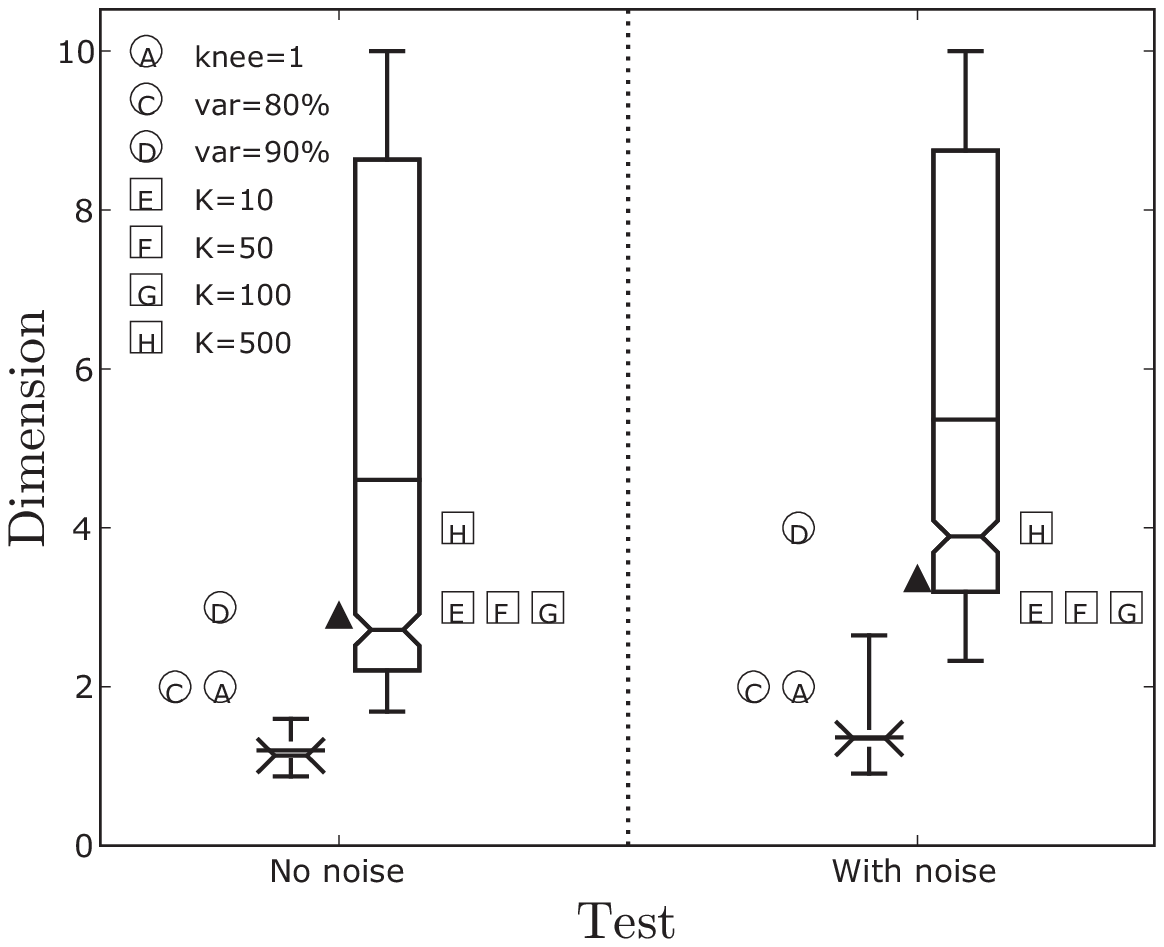} }
\caption[]{\PDE estimates for the data set sampled from a planar disc and spiral embedded in a 5D ambient space.  A description of the box-and-whisker plots is given in Figure~\ref{ball6_results}.
The circular markers indicate the dimension estimation results for PCA using the method indicated.
The square markers indicate the dimension estimation results for Isomap using the parameter indicated. Additional values of $\varepsilon$ were tried that are not marked in this figure, giving estimates of $d=2$ (see Figure~\ref{isomap_discspiral}).
The broad distributions of maximum slopes are due to the data set being heterogeneously distributed in its ambient space (see text for details).
\label{discspiral_dims}}
\end{figure}

In Figures~\ref{discspiral_results} and~\ref{discspiral_dims} a
\PDE analysis of the disc-with-spiral data set indicates the
presence of two manifolds with different dimension, with two
markedly different groupings of slopes of \rV curve corresponding to
$d=1$ and $d=2$. We make four observations about this analysis.

(1) For reference points in the spiral arm, such as those for the cyan
or green \rV curves, the slopes at small radii are approximately one.
Sharp increases in slope are observed at moderate distances
corresponding to balls reaching the central disc or other parts of the
spiral. As such a contact is made a ball will suddenly encounter a
large density of points over a relatively small increase in radius,
and thus the ``volume'' approximated by $\log(k)$ will increase at a
higher rate as $r$ increases. This effect gives rise to secants in the
cyan and green \rV having a spread of maximal slopes $d_x^1 \sim 10$
that do not correspond to the dimension of any manifold. When the
balls increase in radius by the size of the disc's diameter the slopes
in the \rV curves again suddenly flatten out.  For comparison, a
\PDE analysis was repeated on the spiral arm part of the
data set. This produced a scatter plot that only retained the cluster
of points lying along the vertical axis at $d_x^0 \approx 1.05$ (data
not shown), confirming that the disc was entirely responsible for the
slopes clustered at $d_x^1\approx 2$.

(2) For reference points inside the disc, such as that for the yellow
\rV curve, $k(r)\sim r^2$ (indicated by $d_x^1$) until the ball radii
reach the edge of the disc at a radius of 1.8. For the yellow curve with
$x=(-0.186,-0.464)^t$ this happens at approximately $\log(r) = \log
(1.8-|x|) \approx 0.26$.  After the edge is reached, balls grow at a
lower rate, which fluctuates as the balls reach successive parts of
the spiral arm. From the plotted \rV curves, the spread of minimum
slopes $d_x^0$ corresponds to balls centred in the disc having large
enough radii that they are not fully contained inside the disc.

(3) The summary statistics for the \PDE analysis in
Figure~\ref{discspiral_dims} do not adequately represent the bimodal
nature of the distribution of \rV curve slopes shown in
Figure~\ref{discspiral_results}.

(4) The added noise had sufficient amplitude in directions
orthogonal to the plane of the disc and spiral to increase the estimated
dimension by approximately 0.6, as seen in the increase of $\min \:
\dmaxs$ and $\mean \: \dmaxs$ in Figure~\ref{discspiral_dims}.

The results of Isomap for the disc and spiral data set are shown in
Figure~\ref{isomap_discspiral}. In the absence of added noise in the
data sets, Isomap estimated a single two- or three-dimensional
manifold for the disc-with-spiral data set for $K \geq 15$.  Isomap
was very sensitive to the choice of $\varepsilon$, and failed except
at $\varepsilon \approx 0.1$ (0.25) for the noise-free (noisy) data,
when it estimated $d=2$ (3).  The addition of noise to the data set
otherwise made little difference to the estimates of $d$.  The
neighbourhood graphs generated typically contained only one connected
component, but any additional components found contained too few
points to be embedded separately. PCA at 80\%
and 90\% variance capture thresholds esimated $d=2$ for both the
noise-free and noisy data sets, while the estimate from the position
of a knee in the graph of PCA residuals was $d=2$ (3) for the
noise-free (noisy) data.

\subsection{Virtual robot arm data\label{virtualrobot_results}}

\begin{figure}[tbp]
\centerline{\includegraphics[scale=1.1]{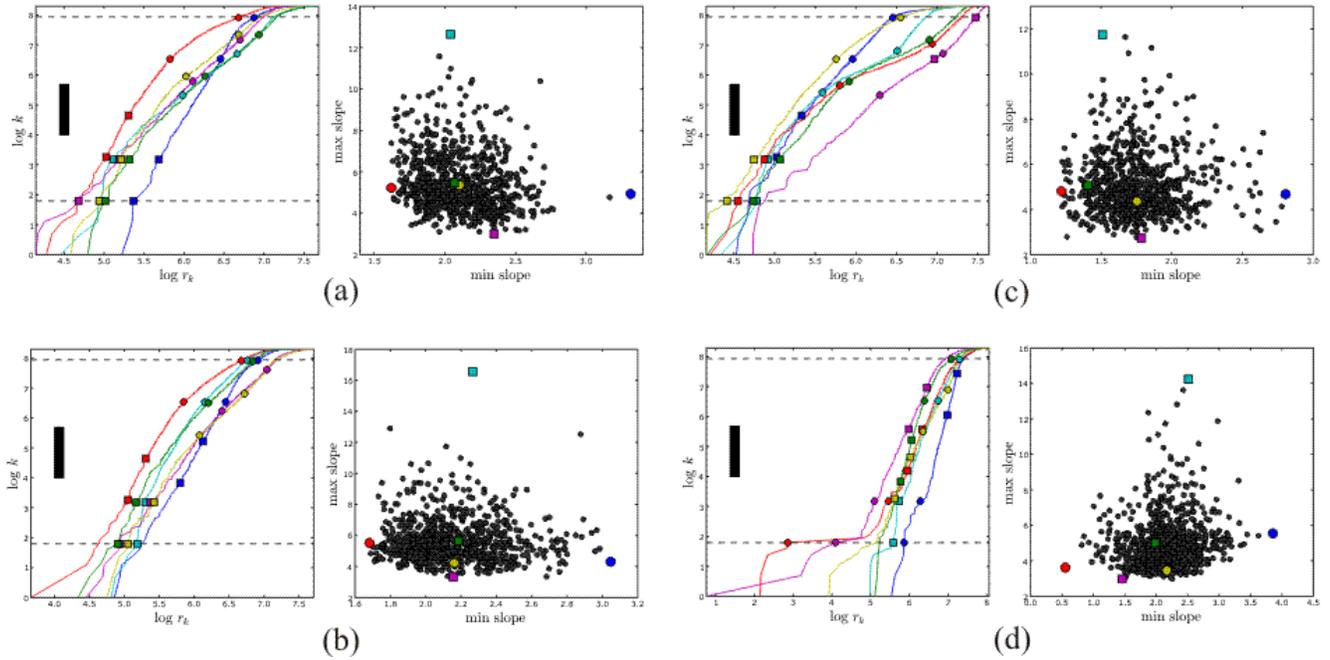} }
\caption[]{\PDE analysis for (a)~the original virtual robot marker placement, and (b)~with more widely spaced markers at joint 6. The position of joint 3 is slaved by that of joint 2 in (c) for the original marker placement. (d)~In contrast to the uniform distribution of joint angles used in (a)--(c), this plot shows the results for joint angles determined by a random walk in joint angle space, using the original marker placement. Almost flat regions of some \rV curves are observed in~(d) for small radii, corresponding to the presence of localized clusters of a few closely-spaced points. The \rV curves in all the plots flatten out slightly at the largest radii, presumably as balls $B_x(r)$ extend outside the data set in some directions but not others.
\label{vr_histos}}
\end{figure}

\begin{figure}[tbp]
\centerline{\includegraphics[scale=0.65]{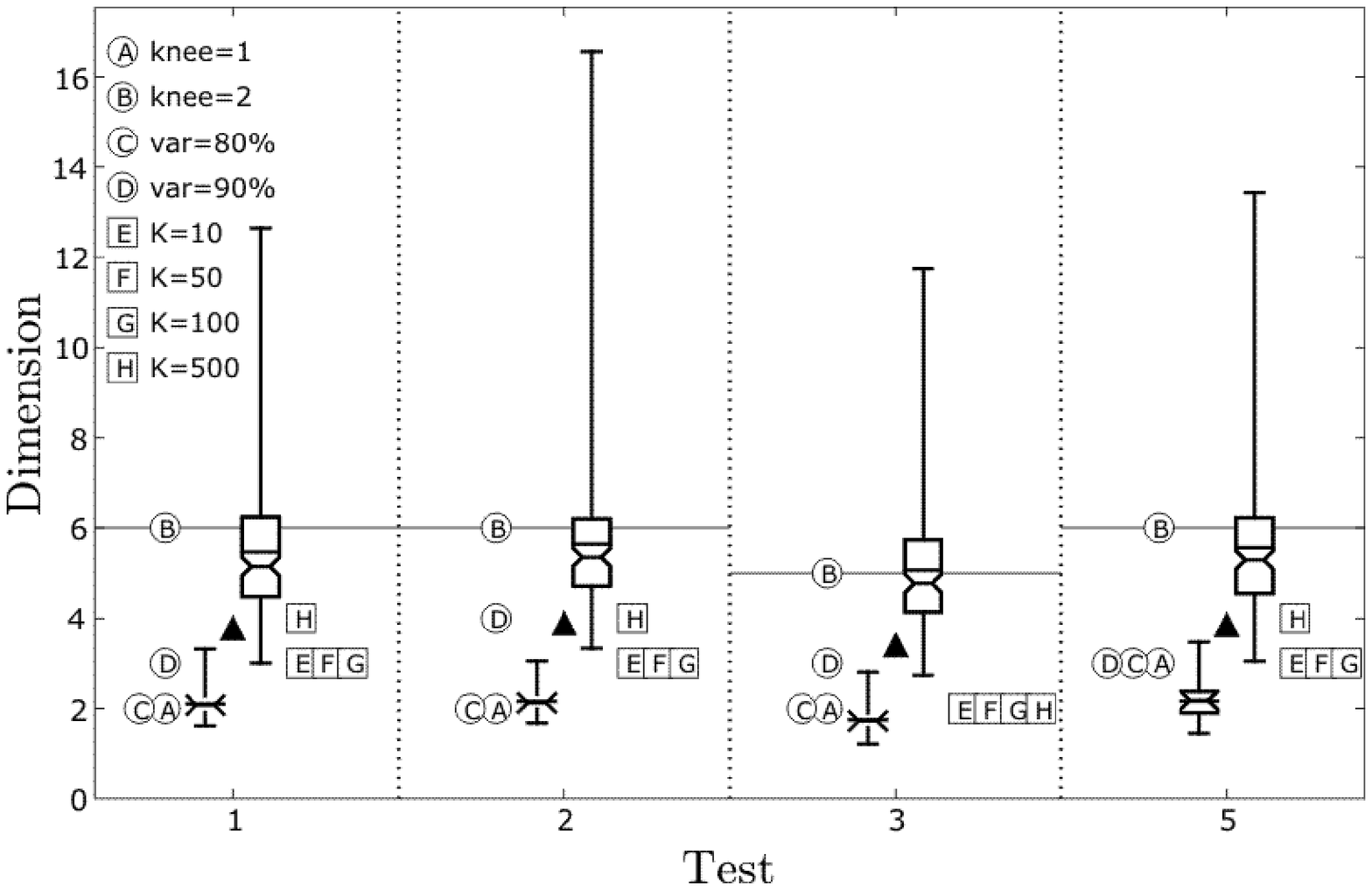} }
\caption[]{Dimension estimates of virtual robot data sets consisting of 4,000 frames. Test 1 uses the original marker placement. Test 2 uses a configuration in which markers were spaced more widely around joint 6. In Test 3, the position of joint 2 is slaved to be a smooth function of the position of joint 3. (Test 4 results are omitted.) Test 5 removes the physically-realistic constraints on joint angles present in the other tests.
A description of the box-and-whisker plots is given in Figure~\ref{ball6_results}.
The circular markers indicate the dimension estimation results for PCA using the method indicated.
The square markers indicate the dimension estimation results for Isomap using the parameter indicated. The results indicate that PCA using a variance capture threshold consistently underestimates $D$.
\PDE, Isomap, and PCA (using residual variances) all succesfully detected that the dimension of the set in Test~3 is one lower than in the other tests. \PDE provides a good estimate near to the mean of the data set for $\dmaxs$. The additional widening of marker placement in Test~2 did not qualitatively affect the estimates, although it made the mean and median $\dmaxs$ values from \PDE more accurate. Test~5 demonstrates that the constraints on arm position make little difference to the dimension estimates.
\label{vr_results}}
\end{figure}

In a marker configuration corresponding closely to that of the
physical robot, 4,000 frames of randomly generated virtual robot arm
postures were analyzed. We first consider the method of joint angle
generation that randomly samples the absolute joint angles from a
uniform distribution. The results of this are presented as Test 1 in
Figure~\ref{vr_results}, and the corresponding \rV curves and scatter
plots from \PDE are shown in Figure~\ref{vr_histos}(a).

\begin{figure}[tbp]
\centerline{\includegraphics[scale=0.45]{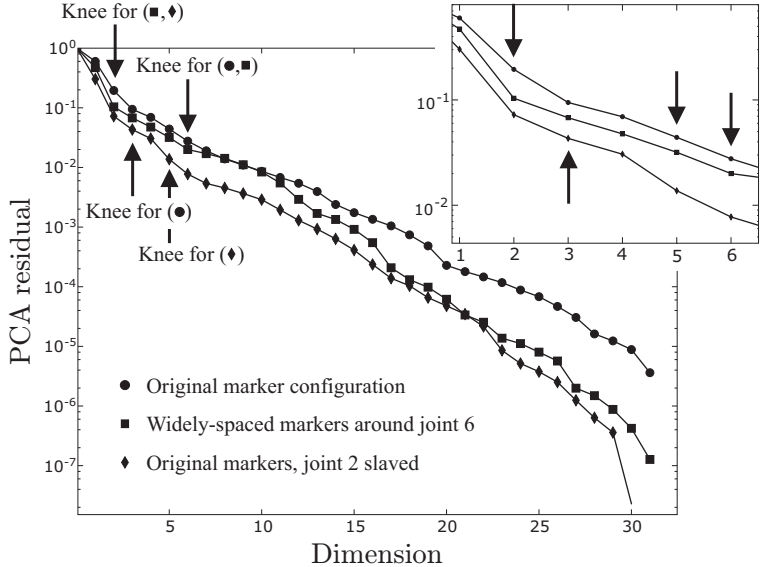} }
\caption[]{Residuals of $d$-dimensional PCA embeddings of the virtual robot motion data. The positions of the first two knees in the graphs are indicated by the arrows. The inset shows the region of the graph around the positions of the second knees.\label{PCAresiduals_vr}}
\end{figure}

The last rigid link in the robot arm is both short (approx.\ 30\,mm)
and narrow (approx.\ 60\,mm). Therefore, the typical distance of the
three markers around the associated joint (\#6) from the joint axis is
small in comparison to those at the other joints.  We investigated the
effect of these multiple spatial scales on the dimension estimation
algorithms by moving the marker positions on the virtual robot further
from the centre of joint 6.  The more widely spaced positions of the
markers around joint 6 were (30, -75, 0), (30, -40, -40), (30, 65, 15)
in the respective local coordinate frames. These are distributed at a
spatial scale comparable to the markers around joint 5. The output of
\PDE for this marker configuration is shown in
Figure~\ref{vr_histos}(b), and the dimension estimate statistics are
presented as Test 2 in Figure~\ref{vr_results}.  Isomap gave similar
results when varying $\varepsilon$ between 80 and 400.  A further
widening of the marker spacing from the joint by 50\% primarily
reduced the inter-quartile range of the \PDE slope estimates (data not
shown). Figure~\ref{PCAresiduals_vr} graphs the residuals of PCA
$d$-dimensional embeddings of three virtual robot motion data sets as
a function of the number of components $d$.

We performed two further tests with the original marker
configuration. In Test 3, we slaved the position of joint 2 to be a
smooth function of the position of joint 3, according to $\theta_2 =
\theta_3^3$. In Test 4, we simply froze the position of joint 3. The
number of DOFs of the system are reduced by one in each case. The PCA
and Isomap results for these two tests are identical, and the
\PDE results for the two tests were almost identical. For this
reason only the statistics for Test 3 are listed in the table.

The virtual setting for the simulated robot arm permitted us to
explore the effect of physical constraints on the dimension analysis
of the physical robot arm's motion. The primary constraints are the
presence of the floor to which the robot is attached and the limited
field of view of the motion capture video cameras.  These constraints
are easy to remove in the simulations, and in Test 5 we generated
4,000 frames of postures for the original marker configuration, in
which joint angles were chosen randomly throughout their full
$360\degree$ range.  (This corresponds to an underlying measure
without boundary.) We observe that the \PDE dimension estimates for
this data set did not change by more than a few percent compared to
Test 2, but the PCA and Isomap algorithms estimated one dimension
greater across the same range of $K$ and $\varepsilon$ values.

\begin{figure}[tbp]
\centerline{\includegraphics[scale=0.65]{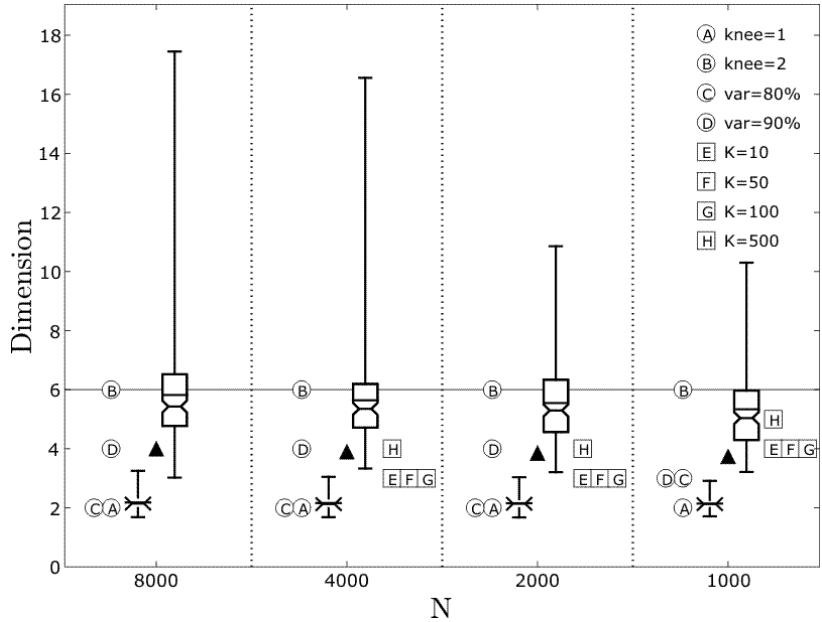} }
\caption[]{Dimension estimates of virtual robot data sets (with more widely spaced markers on joint 6) as $N$, the number of points in the set, is varied.
A description of the box-and-whisker plots is given in Figure~\ref{ball6_results}.
The circular markers indicate the dimension estimation results for PCA using the method indicated.
The square markers indicate the dimension estimation results for Isomap using the parameter indicated.
The mean and median values of $\dmaxs$ in the \PDE distributions become less accurate as $N$ is reduced, but the spread of the distributions also tightens. Isomap also became inaccurate as $N$ decreased to 1,000. PCA results were almost unaffected by the changes in $N$, and the estimation of $D$ using the second knee in the residuals was correct in every case. (Note that Isomap could not be run for $N=8000$ because the data set was too large for the inter-point distance matrix to be calculated.)
\label{vr_results_fewpoints}}
\end{figure}

We studied the dependence of the analyses on the number of points $N$
in the data set by reducing $N$ from 8,000 to 4,000, 2,000, and 1,000,
for the widened configuration of the markers at joint 6 on the
simulated robot.  Figure~\ref{vr_results_fewpoints} presents a
comparsion of the results using the different methods. For this range
of $N$ we did not observe an effect on the position of knees in the
graph of PCA residuals.

We next generated data for the virtual robot arm by performing a
random walk in joint angle space. The results of this are presented in
Figure~\ref{vr_histos}(d).  One major difference in these results is
the presence of \rV curves with very flat regions. These regions cause
the distribution of minimum slopes $d_x^0$ to include values near
zero, and as these regions contain so few points their slopes are not
interpreted as dimension estimates. The initially steep rise of
$\log(k)$ as a function of $r_k$ before a flat region suggests that
there is a small cluster of data points that are very closely spaced
in comparison with the typical distances between other points in the
data set.  The cluster need not be spatially dislocated from the
remaining points, and in this example we verified that the centres of
the dense clusters are indeed located at distances comparable to the
mean interpoint distances between all points in the data set.  Thus,
the presence of the flat regions in these \rV curves may be due to the
fractal structure of random walks and the low dimension of Brownian 
sample paths~\cite{Falconer}.

\subsection{Robot arm motion and the exploration of ``skin'' artifact\label{robot_results}}

\begin{figure}[tbp]
\centerline{\includegraphics[scale=0.45]{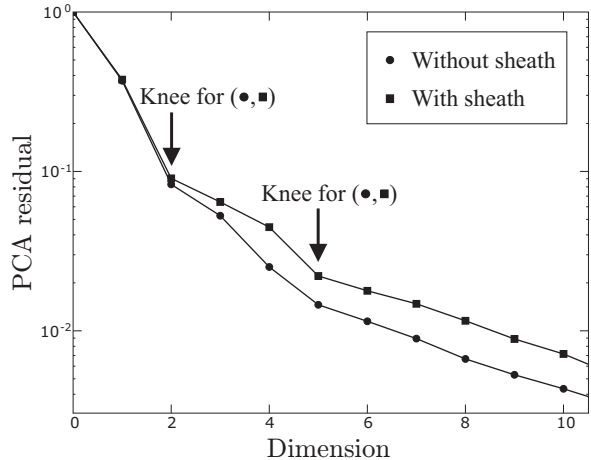} }
\caption[]{Residuals of $d$-dimensional PCA embeddings of the robot motion data.\label{PCAresiduals_robot}}
\end{figure}

\begin{figure}[tbp]
\centerline{\includegraphics[scale=0.95]{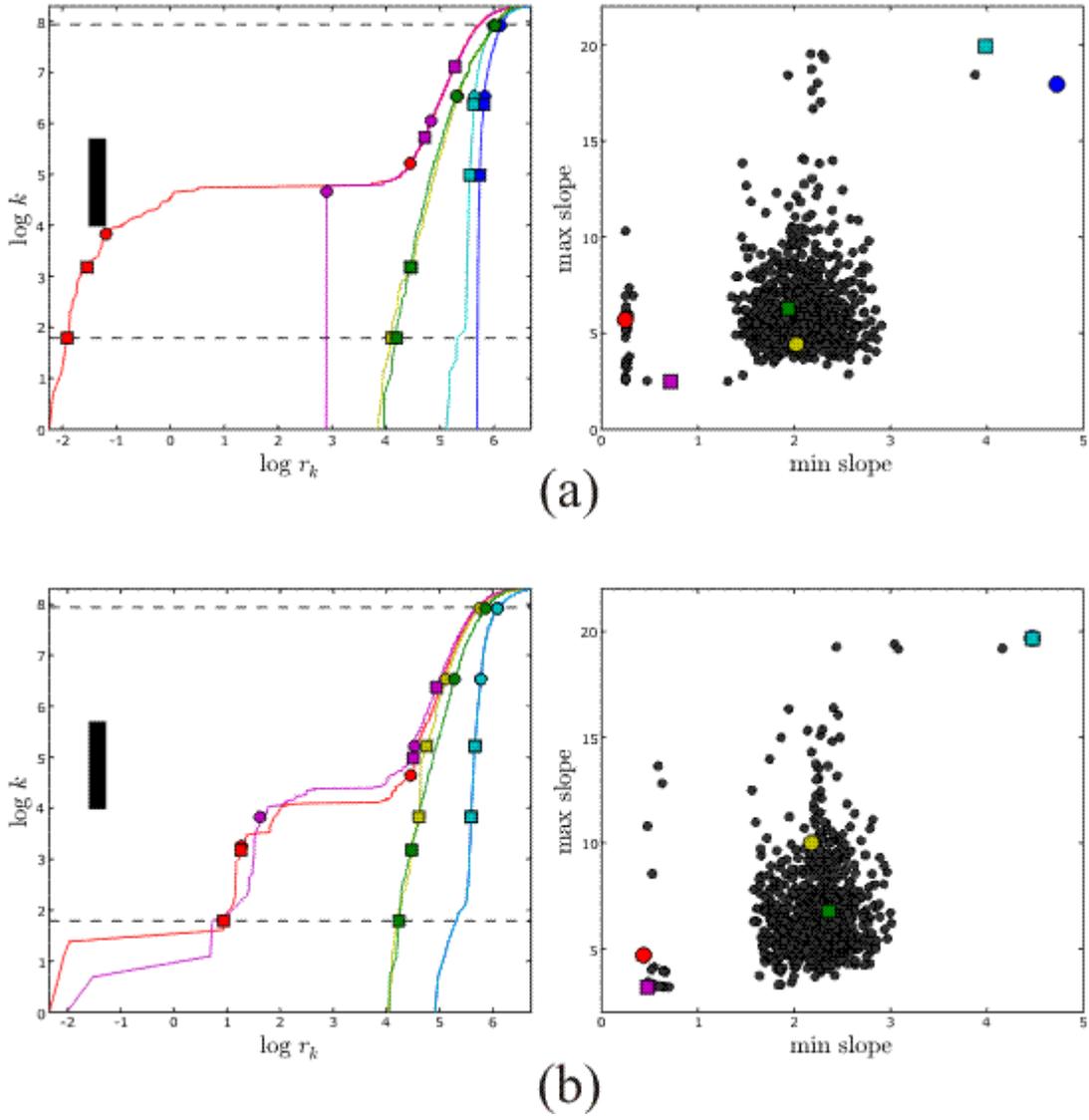} }
\caption[]{\PDE analysis of 4,000 frames of data for (a)~the robot arm without the elastic sheath, and (b)~the robot arm covered with the sheath. Notice that in the scatter plot for (b) the blue circle and cyan square coincide.\label{compare_robot}}
\end{figure}

\begin{figure}[tbp]
\centerline{\includegraphics[scale=0.85]{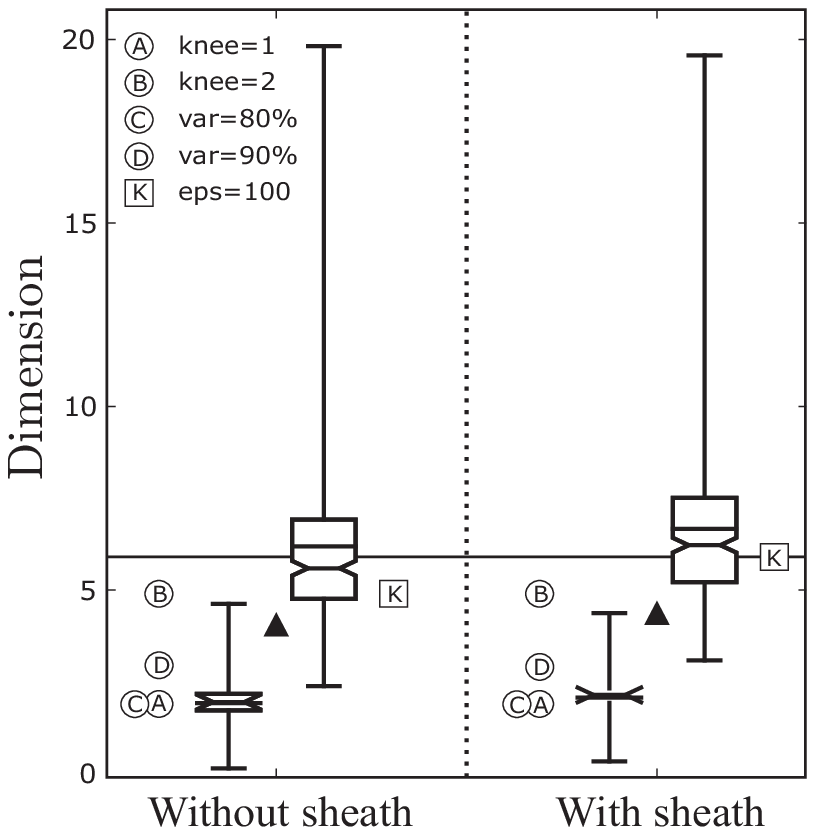} }
\caption[]{PD-E analysis of robot data sets with and without an elastic sheath, using $N=4000$.
A description of the box-and-whisker plots is given in Figure~\ref{ball6_results}.
The circular markers indicate the dimension estimation results for PCA using the method indicated.
The square markers indicate the dimension estimation results for Isomap using the parameter indicated.
The addition of the sheath did not greatly increase the dimension estimates by any of the methods.
\label{robot_spandex_results}}
\end{figure}

Figures~\ref{PCAresiduals_robot}, \ref{compare_robot},
and~\ref{robot_spandex_results} show the results of our analysis of the
robot motion capture data. The mean values of the maximum and minimum
slopes are close to those predicted from the virtual robot tests, and
the qualitative pattern of point distribution in the scatter plots is
similar to that in the virtual robot results when the method of joint
angle generation is the same (see Figure~\ref{vr_histos}(d)).
The addition of an elastic sheath to the robot did not significantly
change the dimension estimates, and in particular made almost no
difference to the inter-quartile ranges of $\dmins$ or $\dmaxs$.

\subsection{Hand motion\label{hand_results}}

\begin{figure}[tbp]
\centerline{\includegraphics[scale=0.6]{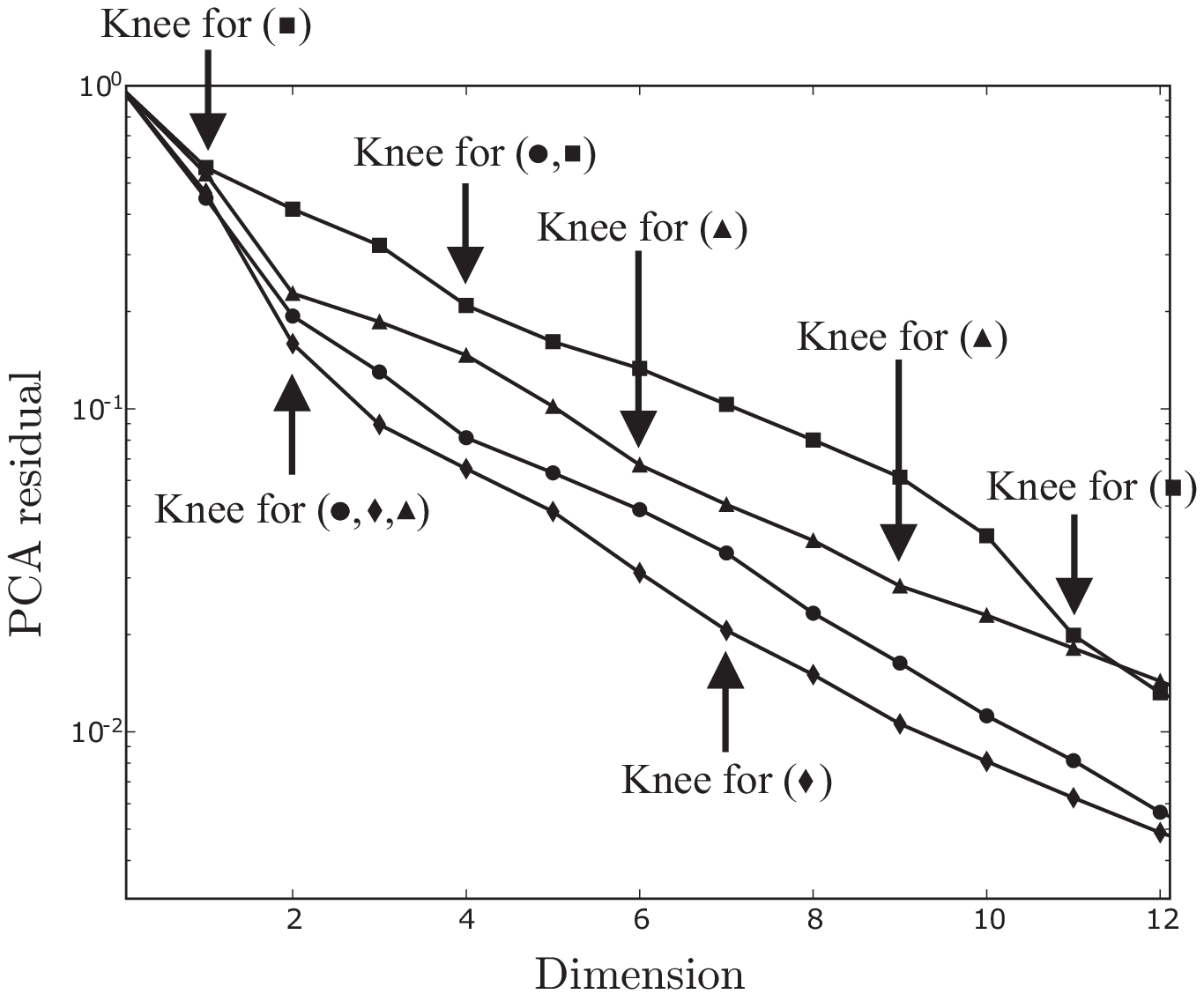} }
\caption[]{PCA residuals for all hand motion tasks, showing first two knees for each task. Random motion task (subject S2): circle markers; (subject S1): square markers. Keyboard task (subject S2): triangle markers. Trackball task (subject S1): diamond markers.\label{hand_PCAresiduals}}
\end{figure}

\begin{figure}[tbp]
\centerline{\includegraphics[scale=1.1]{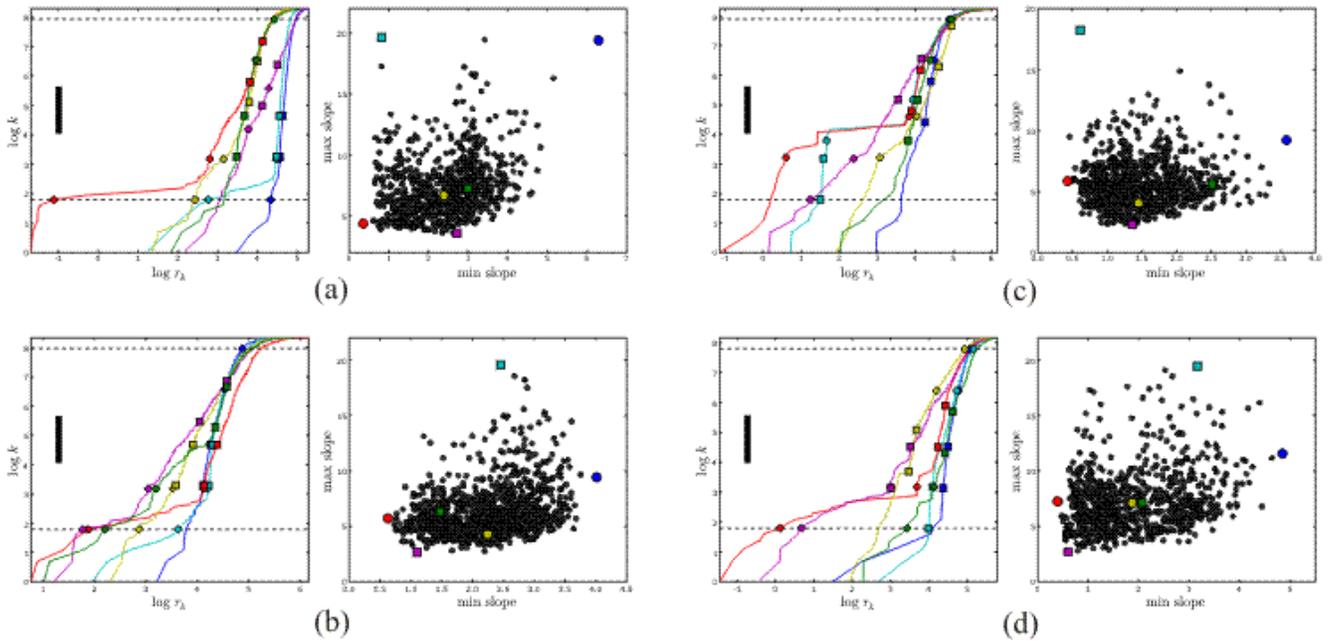} }
\caption[]{\PDE analysis for 4,000 frames of hand motion data. (a)~Random motion task (subject S1).(b)~Random motion task (subject S2). (c)~Trackball simulation task (subject S1). (d)~Keyboard simulation task (subject S2).
Notice the presence of flat regions in the some \rV curves for every task, and the mild flattening of the curves for radii $r > 5$.
\label{PDE_hand_results}}
\end{figure}

\begin{figure}[tbp]
\centerline{\includegraphics[scale=0.35]{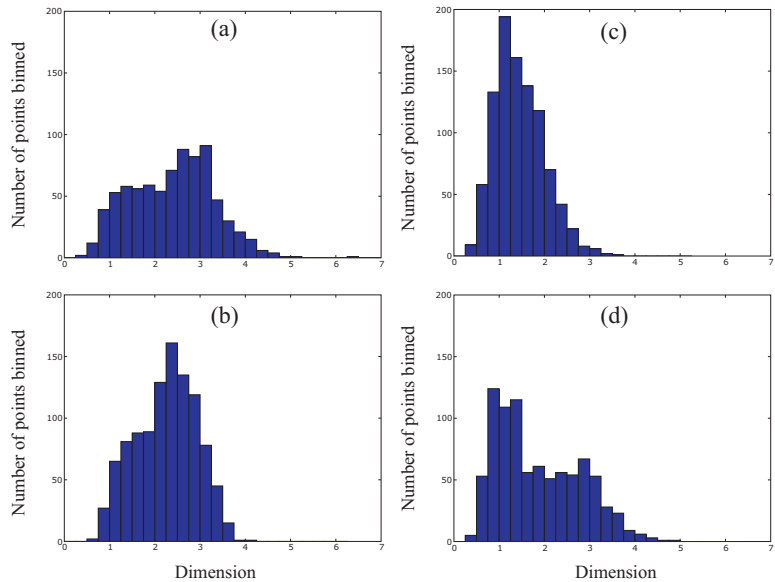} }
\caption[]{Histograms of minimum slopes $d_x^0$ with bin width 0.25 for the hand motion tasks. (a) Random motion task (subject S2); (b) Random motion task (subject S1); (c) Keyboard task (subject S2); (d) Trackball task (subject S1).
These histograms show the roughly bimodal nature of the minimum slopes for all but the trackball task in~(c), which appears to contain only a component centred at $d \approx 1$.
\label{Hand_histos}}
\end{figure}

\begin{figure}[tbp]
\centerline{\includegraphics[scale=0.7]{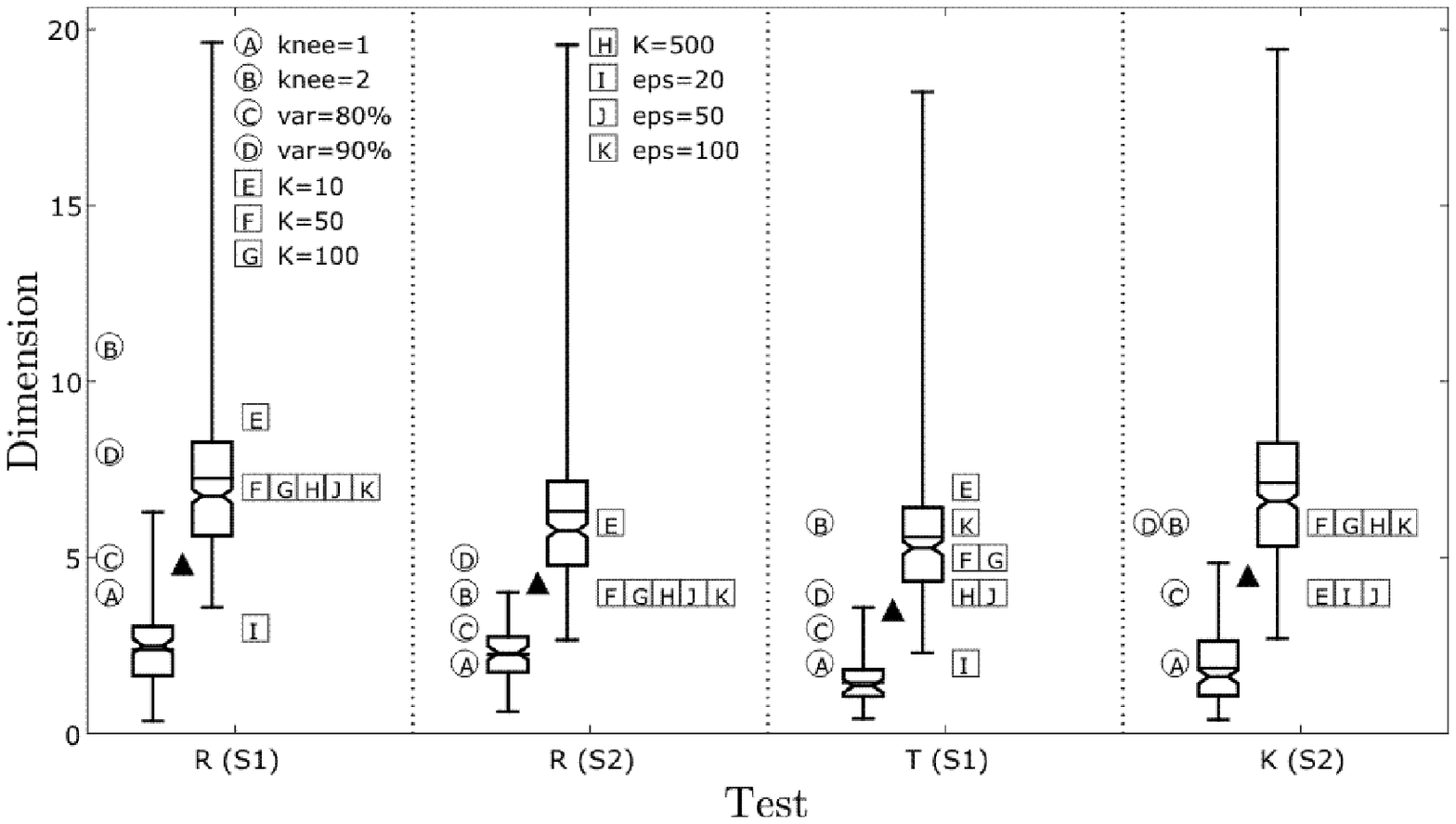} }
\caption[]{Dimension estimates of hand data sets from a sampling of 4,000 points. Task {\em R}: Random motion. Task {\em T}: Simulation of trackball manipulation. Task {\em K}: Simulation of computer keyboard use.
A description of the box-and-whisker plots is given in Figure~\ref{ball6_results}.
The circular markers indicate the dimension estimation results for PCA using the method indicated.
The square markers indicate the dimension estimation results for Isomap using the parameter indicated.
Qualitatively, both PCA and Isomap predict estimates less than $d=10$. Most of the \PDE estimates are also in this range, although there are a few maximum slopes at higher $d$ values. There is not a clear quantitative trend in the estimates as a function of the task in this preliminary study using only two subjects.
\label{table_hand_results}}
\end{figure}

Our analysis of four data sets for human hand motion is given in
Figures~\ref{PDE_hand_results} and~\ref{table_hand_results}.  The
PCA residuals for these data sets are plotted in
Figure~\ref{hand_PCAresiduals}. The most evident result is that all
the methods estimate the dimension of hand motion to be less than 11,
and probably around 6. Also, a histogram of the minimum slopes $d_x^0$
for the trackball task indicates the absence of a dense cluster at
$d_x^0 \approx 3$ in contrast to the other data sets
(Figure~\ref{Hand_histos}). This may indicate that the appropriate
dimension reduction for hand motion may be task dependent, but requires
a more systematic analysis with more data.

In the scatter plots we observe a large number of points for which $\min \:
\dmins \approx 0.5$. The red \rV curves corresponding to the minimum
$d_x^0$ secant slopes indicate that these are due to flat regions of
the curve.  In the study of the synthetic data sets we previously
identified such flat regions as corresponding to localized subsets of
a small number of closely-spaced points. In this case there appear to
be many such small subsets. (Recall that these flat regions are due to
this localization and that the associated low slopes are not
indications of low dimension.) It will be instructive in future work
to identify which hand postures are associated with these subsets.

\section{Discussion\label{discussion}}

Motion capture data produces measurements of marked positions of an
object in a $D$-dimensional data space. The number of kinematic
degrees of freedom (DOFs) of the object is often substantially smaller
than $D$, and the system may be constrained to a set of even smaller
dimension than those which are kinematically feasible. In response to
a growing trend in biomechanical and neurophysiological analysis to
estimate the number of DOFs of a system (i.e., its ``dimension'')
using linear methods such as PCA, we compared PCA to two nonlinear
methods (Isomap and the novel pointwise dimension estimation (\PDE)
algorithm) that are also designed for this purpose.  We compared and
contrasted the dimension predictions by the three methods using
simulated kinematic data, and motion capture data from a robot arm and
from human hands whose degrees of freedom are known and unknown,
respectively. Additionally, we challenged the idea that a method such
as \PDE, that is based on calculations of a type of fractal dimension,
requires unreasonably large amounts of data.

We now summarize conclusions about the three methods of dimension
estimation, followed by a discussion of the results for the tests on
hand motion.

\subsection{Interpretation of the pointwise dimension results}

The \PDE algorithm assumes that the data set is a discrete approximation
of a probability distribution and seeks asymptotic power law
relationships of the form $V \sim r^d$ for the volumes $V$ of balls as
measured by this distribution and their radii $r$.
The log-log plots of $V(r)$ vs. $r$, which we refer
to as \rV curves, show this relationship for balls with a common
center. The calculation of these curves is easy, but deriving an
estimate for $D$ from these curves encounters several issues that
complicate the analysis.  Statistical sampling fluctuations and noise
make estimates of $V(r)$ unreliable for small values of $r$, and the
scaling relationship may break down for large values of
$r$. Therefore, implementation of the methods seeks an intermediate
``scaling region'' in which the \rV curves are approximately linear in
log-log plots. At present there is not a complete analytical framework
for applying the definition of pointwise dimension to experimental
data in this way. As a first step, therefore, we have adopted an
empirical approach to testing whether our \PDE algorithm provides
reasonable estimates with data sets of a few thousand points
approximating measures of known dimension.

The above complications to reliably estimating the volumes of balls
using \rV curves mean that \PDE analysis results in a statistical
distribution of dimension estimates. This distribution requires
interpretation by the user in the context of the experimental system
and the quality and quantity of data. Primarily, the distribution
provides an upper bound on the estimated dimension, and sometimes a
lower bound if the lower end of the distribution tails off at a value
greater than 1.

For smooth measures supported on manifolds with boundary we have found
that the method produces a distribution of slopes for the
\rV curves whose upper bound is close to the known pointwise
dimension of the measure. Typically, the slopes of the \rV curves are
smaller than the pointwise dimension of the underlying measure due to
``boundary effects.''  As the dimension of the manifold grows, an
increasing proportion of points of the manifold lie near the boundary.
Increasing dimension limits the radii of balls that do not intersect
the manifold boundary. Simultaneously, statistical sampling
fluctuations for balls of a fixed radius increase.

\begin{figure}[tbp]
\centerline{\includegraphics[scale=0.45]{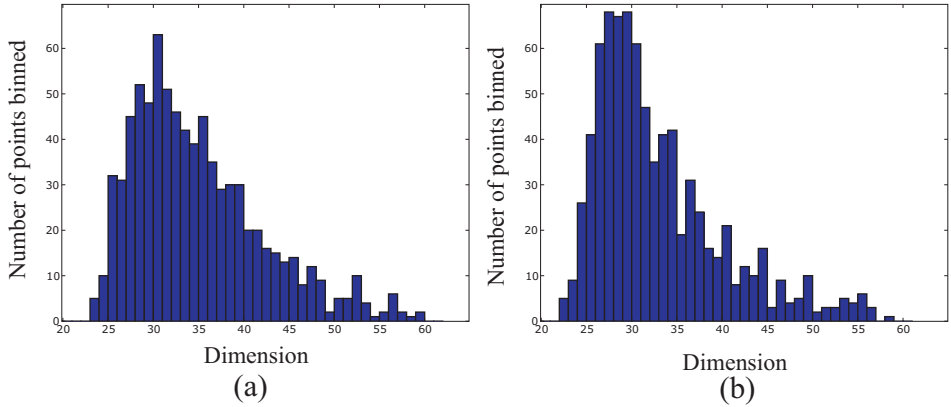} }
\caption[]{Histograms of maximum slopes $d_x^1$ using a unit bin width
for a 54-dimensional rectangular solid (a) with equal side lengths,
and (b) with 4 sides one fifth the length of the other
50. These show a concentration of slopes near $D/2 = 27$ and a long tail of relatively
few binned points from the scatter plots in Figure~\ref{random_rect_histos_54D}.
\label{rect54_histos}}
\end{figure}

For measures that were intrinsically high dimensional (such as
independent samples from rectangular solids), the distribution of
slopes measured from the \rV curves tends to cluster at values
intermediate between $D/2$ and $D$. (For instance, this can be seen in
the histograms of binned $d_x^1$ values for the 54-dimensional
rectangular solids, in Figure~\ref{rect54_histos}.) The mathematical
analysis of the disparity between $D$ and the slopes of the \rV curves
warrants further study.

However, in the tests on low-dimensional data
sets of known dimension, one or other of the median values of $\dmins$
or $\dmaxs$ gave a good estimate of the dimension. For instance, the
median of $\dmins$ was a good estimate for the dimension for the curve
embedded in a Swiss roll manifold, and for the spiral arms of the 5D
data set also including a 2D disc. The median of $\dmaxs$ was a good
estimate for the disc of this latter data set, as well as for the 6D
solids and both the real and virtual robot motion capture data. Note
that in all cases direct inspection of the \rV curves greatly aids in
the interpretation of the slope statistics.

For the sets of unknown dimension (the AdeptSix 300 robot arm and the
human hands), PCA and Isomap both predicted $d$ values much less than
the maximum of $\dmaxs$ from the \PDE analysis. From our experience
with the data sets of known dimension, these $d$ values are in a low
range for which the mean and median values of the \PDE slopes for
$\dmins$ or $\dmaxs$ also predicted $d$ accurately. We conjecture that
this will also be the case for the robot arm and human hands, so that
our conservative estimate using \PDE need not extend to $d=20$ at the
maximum value of $\dmaxs$, but instead could be set at $d=10$.

The \PDE method also appears to be sensitive to data sets
that can be partitioned into subsets having different dimension, such
as the union of a disc and a spiral. The partitioning appears as
clustering in the scatter plots and a bimodal distribution of slopes
for the \rV curves.

We speculate that the process of generating postures in animal motion
may be directly responsible for the presence of the flat regions in
the \rV curves and the associated clusters of closely-spaced points.
We observe that typical hand motions for the tasks in this study
consist of a sequence of somewhat discrete changes in individual
finger positions or in coordinated groupings of fingers (e.g., flexing
multiple fingers simultaneously), while other parts of the hand may
remain stationary. Sampling from such motion may generate sequences of
postures not unlike those generated by a random walk, where almost all
sample paths have dimension $2$, independent of the dimension $(\ge 2)$
of the ambient space in which the walk takes place. We observed earlier that
postures generated by random walks in our benchmark tests created
localized subsets of closely-spaced points in the data, characterized
by \rV curves having regions that are almost flat. These flat regions
were not observed when postures were uniformly sampled. We have not 
explored how long we would need to observe our robot before its observed 
random path would be expected to fill the six dimensional rectangular solid
in joint space uniformly.

Our implementation of the \PDE algorithm is suitable for data sets of
a few thousand points. The algorithm can analyze larger data sets than
similar algorithms that calculate a complete matrix of interpoint
distances for points in a data set.  Our results were insensitive to
the size of data sets in the range 1,000--8,000. However, experimental
trials as long as possible are still necessary (for any method) to
ensure that the sampling of postures from the robot or hand (a)
approximate the full distribution of postures assumed in the specified
task; and (b) are at a sufficiently low sampling rate to prevent close
correlations between successive samples.  The analysis summarized in
Figure~\ref{vr_results_fewpoints} showed that we required fewer data
points than were collected in order to achieve robust results from
\PDE.

We conclude from these tests that an algorithm such as the
\PDE method presented here is effective in the {\em qualitative}
exploration of complex geometric structure in motion capture data, and
directly estimates bounds on the dimension of the set of postures
assumed by an object in data space. Although this estimate is likely
to be systematically biased toward an underestimate of the dimension
we believe that it is feasible to correct this bias: these
underestimates could be quantified by further empirical studies using
benchmark data sets and their relationship to $D$ analyzed.
Furthermore, the lack of differences between the estimated DOFs of the
virtual and real robots suggest that we can expect improvement in the
accuracy of dimension estimation for experimental data sets by
continuing to refine our methods using abstract test data. In future
work our method would be augmented by the parameterization of
low-dimensional manifolds identified by nonlinear dimension estimation
methods using PCA or more general techniques (e.g., kernel
PCA~\cite{kernelPCA}), which would elucidate finer detail in the
geometric structure of the manifolds and their orientation in the
whole data set.

\subsection{Interpretation of Isomap results}

The primary control parameters for Isomap are the neighbourhood size
$K$ or the neighbourhood radius $\varepsilon$ and the number of
coordinates used to represent a manifold. In the Swiss roll example
there is a clear separation of spatial scales, and $\varepsilon$ is
easily tuned to a scale much smaller than the separation of the folded
layers of the manifold so that Isomap detects the one-dimensional
nature of the curve rather than the globally three-dimensional nature
of the whole set of points in the embedding space. However, in our
motion capture data of a robot arm and a human hand there are no
strongly separated spatial scales. The choice of appropriate values of
$K$ or $\varepsilon$ for Isomap is more problematic in these cases,
although \PDE analysis of the data sets can be used to predict good
starting choices for these parameters. For instance, in the robot arm
tests we obtained the best Isomap results with $\varepsilon=100$.  In
the \PDE analysis of the same data set we observed $\left[ 90, 150
\right]$ to be the approximate range of ball radii corresponding to
points near the centre of the largest cluster in the scatter plot
(Figure~\ref{compare_robot}).  Similarly, the characteristic number of
points in those balls seem to provide good starting choices for the
Isomap parameter $K$. For the robot arm this corresponds to $K\approx
50$.

\subsection{Interpretation of PCA results}

PCA is frequently used as a technique for dimension reduction, seeking
to minimize the variance of residuals among projections of data in
$\mathbb{R}^D$ onto a $d$-dimensional (linear) subspace. If this
projection preserves relevant properties of the data set, then $d$ is
regarded as an (upper) estimate for the dimension of the data
set. Here we are primarily interested in methods that choose minimal
values of $d$ that appear to preserve the geometric properties of the
set, because these properties inform the process of modelling a
biomechanical system as a dynamical system. The circumstances in which
PCA works well to do this are ones in which the data is composed of a
low dimensional signal and noise that is largely orthogonal to the
subspace containing the signal. In these circumstances PCA will retain
the signal while removing much of the noise from the data. It is
hardly apparent that the underlying geometric structures produced by
biomechanical systems have these properties, so we do not presume that
PCA will be the most effective estimator of the DOFs of those systems.

PCA is commonly used to estimate dimension based on a fixed variance
capture threshold $\tau$. However, the presence of nonlinear geometric
structure in high-dimensional data sets means that the resulting
estimates will be sensitive to noise or other small changes in the
data if $\tau$ is not chosen to suit the properties of the data.
Large values of $d$ occur when there are large numbers of small singular
values. This implies that the cumulative norm of the first $k$
singular values changes slowly with $k$ as $k$ grows large, making the
choice of $d$ highly dependent upon the choice of $\tau$.

The minimum value of $\tau$ needed to obtain a dimension estimate of
$d$ for a data set with dimension $D \geq d$ is defined by the maximum
possible value of $\sigma(d)$. This maximum is $\sqrt{d/D}$ for
any $d$, and is obtained when all the singular values of the PCA
decomposition are equal (for instance, in the case of a
$D$-dimensional ball).

We find that graphing the PCA residuals $\rho(k)$ allows a definition
of this threshold in terms of ``knees'' that appear to be better tuned
to the properties of high dimensional data sets. In our exploration of
synthetic and motion capture data sets we found that this method led
to a more consistent and accurate upper estimate for the dimension of
the data sets than by {\em a priori} selection of variance capture
thresholds.

\subsection{Significance for biomechanics\label{biomech}}

Biomechanical systems typically exhibit complex behaviour in their
motion at multiple scales of time and space.  As a result, the
observable degrees of freedom expressed by the system can depend on
the measurement scales chosen by the observer. As a prelude to
developing dynamical models for the motion of a biomechanical system,
we would like to characterize the number of ``active'' DOFs of the
system to which the neuro\-muscular control has access. This number
would be the most useful guide to the number of variables needed in a
dynamical model. However, our measurements are complicated by the
effects of instrumentation noise, and more importantly by ``passive''
DOFs of the system that are inherent in its mechanical
properties---for instance, the elasticity of the skin surface on which
we place reflective markers.  In this paper we have begun to explore
the effects on dimension estimation by systematic noise such as the
passive properties of a skin-like barrier between the motion of a
rigid body and the measurement instruments.

The measurement of the active DOFs is further complicated by the
presence of correlations between observable variables, which we refer
to as ``synergies''~\cite{dAvella}. The nervous system is involved
with monitoring and controlling possibly many thousands of DOFs as
part of its motor control functions. For this reason it is believed to
utilize coordinated patterns of motor activity that act
synergistically on multiple DOFs at once when performing different
tasks, possibly taking advantage of basic biomechanical properties of
the skeletomotor apparatus~\cite{Tresch_matrix,Tresch_motor}.
Therefore, we expect the dynamics of neuromuscular control systems to
be constrained to subsets of relatively low dimension when
performing complex motions such as locomotion or manipulation.

Our analyses with \PDE and Isomap demonstrate the existence of
low-dimensional nonlinear structures in constrained hand motions, such
as those involving rhythmic motion retracing a closed curve in state
space. In these cases PCA typically overestimates the DOFs involved
because PCA is insensitive to nonlinear structure. This is apparent
from the results for the synthetic test data sets consisting of the
curve on a Swiss roll manifold and the union of a disc and a spiral
embedded in a 5D ambient space. In the latter example, Isomap also
fails to detect the one-dimensional nature of the spiral arms because
the method is less sensitive to non-smooth data sets consisting of
distinct subsets having different dimension.

Using \PDE we placed a conservative upper bound on the dimension of
such constrained hand motions at roughly 10.  However, we observed a
high frequency of slopes in the range of 2--3 and 5--10 in the \rV
curves, estimates of 1--2 and 4--7 from PCA when detecting knees in
the residuals, and estimates mostly in the range of 4--7 from
Isomap. Thus, the combined estimates from the three methods we have
focused on suggest that the neuromuscular control of hand motion
involves a similar number of DOFs to that estimated by Santello et
al.~\cite{PCAhand}. In that study the DOFs of hand motion were
estimated in the context of a data set consisting of a wide range of
static hand postures for grasping objects using two methods: PCA
yielded an estimate of approximately 3 DOFs, and a combination of
discrimant analysis and information theory estimated an upper bound of
5 or 6 DOFs. The {\em quantitative comparison} between the
predictions of each method provide a less consistent picture of the
differences between each task, but we plan to elucidate these
comparisons further in future work that uses more subjects.

We noted the possible presence of multiple modes in the distributions
of the slopes of \rV curves for hand motion. This suggests that the
underlying sets of postures in data space might not be smooth
manifolds, and might instead be partitioned into subsets of different
dimension.  One hypothesis to explain this would be that hand motion
involves motor subsystems expressing different numbers of DOFs
depending on context. Testing this hypotheses will therefore require
nonlinear methods such as those discussed here. We plan to
systematically explore this possibility in future work, both in terms
of the dimension of hand motion capture data and of corresponding
electromyographic data from hand muscles.

\section*{Acknowledgment}

This material is based upon work supported by the National Science
Foundation under Grant No.\ 0237258 and FIBR Grant No.\ 0425878;
This publication was made possible by Grants Nos.\ AR050520 and
AR052345 from the National Institutes of Health (NIH). Its contents
are solely the responsibility of the authors and do not necessarily
represent the official views of the National Institute of Arthritis
and Musculoskeletal and Skin Diseases (NIAMS), or the NIH.

\bibliography{Clewley_etal_PD-E}

\end{document}